\documentclass[aps,pre,twocolumn,groupedaddress,showkeys,superscriptaddress]{revtex4-1}
\usepackage{graphicx}% Include figure files
\usepackage{dcolumn}% Align table columns on decimal point
\usepackage{bm}% bold math
\usepackage{epstopdf}
\usepackage{amsmath}
   \usepackage[all]{xy}
\usepackage{subfigure}
\usepackage{color}
\newcommand\wh[1]{\widehat{#1} }
\begin{document}

\preprint{1}

\title{Nonequilibrium fluctuation response relation in a time scale separated system}
 
 \author{Shou-Wen Wang}
 \affiliation{Beijing Computational Science Research Center, Beijing, 100094, China}
 \affiliation{Department of Engineering Physics, Tsinghua University, Beijing, 100086, China}
\author{Kyogo Kawaguchi}
\affiliation{Department of Systems Biology, Harvard Medical School, Boston, MA 02115, USA}
\author{Shin-ichi Sasa}
\affiliation{Department of Physics, Kyoto University, Kyoto 606-8502, Japan}
\author{Lei-Han Tang}
\affiliation{Beijing Computational Science Research Center, Beijing, 100094, China}
\affiliation{Department of Physics and Institute of Computational and Theoretical Studies, Hong Kong Baptist University, Hong Kong, China}

\date{\today}
  \graphicspath{{./figure/}}


\begin{abstract}
We present a theoretical framework to  analyze  the violation of  \emph{fluctuation-response relation} (FRR) for  any observable from a finite Markov system with two well-separated time scales.    We find that, generally for both slow and fast observables,   a broad plateau exists in the intermediate frequency region,  which contributes to a finite hidden entropy production.    Assuming that  non-equilibrium behavior arises only from coupling of slow and fast processes,  we find that, at large observation time scae,  the effective temperature for a slow observable deviates only slightly from the bath temperature, accompanied by an emerging well-defined effective potential landscape,    while the deviation is significant for a fast observable.   Our study also identifies a wider range of applicability of the Harada-Sasa equality in Markov jumping systems.  
\end{abstract}

\maketitle
\section{Introduction}

In the past  three decades, much efforts have been devoted to understanding non-equilibrium thermodynamics,  focusing especially on mesoscopic scale non-equilibrium phenomena,  where fluctuations are strong~\cite{sekimoto2010stochastic,seifert2012stochastic}.  These theoretical studies have led to interesting applications in biological systems, including   kinetical proof-reading~\cite{hopfield1974kinetic,Sartori2013Kinetic}, environment sensing and adaptation~\cite{lan2012energy,mehta2012energetic,Stefano2015Information},  and oscillation maintenance  within cells~\cite{cao2015oscillation},      efficiency  of nanosized rotary motors~\cite{noji1997direct,toyabe2011thermodynamic}.   Due to the intrinsic complexity of these molecular devices and machines,  coarse-graining is inevitable to simplify the description.

Recently,  there have been  much interest to consider the effect of coarse-graining on entropy production of the system~\cite{rahav2007fluctuation,puglisi2010entropy,celani2012anomalous,bo2014entropy,nakayama2015invariance,Chun2015fast,Esposito2015Stochastic,shouwen2015adaptation,chen2016Model},  which is a key thermodynamic quantity that measures how far the system is away from equilibrium.     For a system with two distinct time scales,  coarse-graining is achieved by adiabatically eliminating the fast variables to obtain a simpler description in terms of the slow variables.    Hondou \emph{et al.} first realized that,  for a Brownian particle moving in a spatially modulated temperature field,   over-damped approximation gives the wrong entropy production rate compared with the underdamped Langevin description,  which implies that Carnot efficiency cannot be achieved  in such a Brownian heat engine~\cite{Derenyi1999efficiency,hondou2000unattainability}.   Esposito made a systemmatic study on Markov systems with two time scales and found that  the total entropy production can be decomposed into that  at the coarse-grained level,  that only due to the fast dynamics, and that due to the coupling between slow and fast dynamics.  Surprisingly,  the  coupling term is non-negative and thus cannot be ignored in general~\cite{esposito2012stochastic}.   The missing contribution at the coarse-grained level is referred to as ``hidden entropy production" by Kawaguchi \emph{et al.},  who further showed that it satisfies fluctuation theorem~\cite{kawaguchi2013fluctuation}.  In our very recent paper,  we showed that hidden entropy production reveals itself as a characteristic plateau in the  violation spectrum of \emph{fluctuation-response relation} (FRR) of a slow observable~\cite{Wang2016entropy}.  Our discovery suggests a way to reveal the hidden fast dissipative processes by just studying the trajectory of a slow variable with sufficient temporal resolution.

FRR is a  fundamental property of  an equilibrium system~\cite{kubo1966fluctuation},  and its violation can be exploited  to characterize non-equilibrium  system.  This has been applied to study  active hair bundles~\cite{martin2001comparison},   active cytoskeletal networks~\cite{Mizuno2007cytoskeletonNetwork},  and molecular motors~\cite{toyabe2010nonequilibrium}.   Furthermore, one may  introduce effective temperature as  the ratio between correlation and response,  and use  its deviation from bath temperature to quantify deviation of the system  from equilibrium.   Although initially proposed by  Cugliandolo \emph{et al.}  to characterize glassy systems~\cite{LeticiaPREtemperature}, it has been used recently for small molecules driven out of equilibrium~\cite{dieterich2015single}.  A more foundamental connection between FRR violation and dissipation   was pointed out by Harada and Sasa a decade ago in the context of general Langevin systems~\cite{harada2005equality,harada2006energy}.  Referred to as  the Harada-Sasa equality,  this relation has been confirmed experimentally in a driven colloid system,  and has also been used to study the energetics of   F1 ATPase~\cite{toyabe2007experimental},  a rotary motor with much higher complexity~\cite{toyabe2010nonequilibrium}.  Although there is no general connection between FRR violation and dissipation in discrete Markov systems,   Lippiello \emph{et al.} generalized the Harada-Sasa equality to \emph{diffusive} Markov jumping systems where the entropy production in the medium for each jump is relatively small~\cite{Lippiello2014fluctuation}.  However,  the requirement for this generalization is still not very clear.

 In this paper,  we systematically discuss the FRR violation of a Markov system with two distinct time scales and a finite state space.   Its FRR violation spectrum for both a slow  and a fast observable is derived,  and the connection to hidden entropy production and effective temperature is also discussed.   The paper is organized as follows.  Section~\ref{sect:HS} gives a brief introduction to  Harada-Sasa equality.  Its generalization to  Markov jumping systems is discussed in Section~\ref{sect:generalization_HS}.  In Section~\ref{sect:CR}, we derive  the analytical forms of correlation and response spectrum for Markov systems.    Section~\ref{sect:time scale} introduces our finite Markov model with two  time scales,  followed by a perturbative analysis of this system,   and its FRR violation spectrum for fast and slow observables,  respectively.  In Section~\ref{sect:discussion},  the connection to entropy production  partition and to effective temperature are discussed.   Section~\ref{sect:adaptation} illustrates our main idea through an example of sensory adaptation in E.coli.  We conclude in Section~\ref{sect:conclusion}.

\section{The Harada-Sasa equality}
\label{sect:HS}
The FRR violation of a specific degree of freedom can be related to the dissipation rate of the same degree of freedom,  as examplified by the Harada-Sasa equality   in the context of Langevin systems.   Consider the following $N_0$-component Langevin equation
\begin{equation}
\gamma_j \dot{x}_j=F_j(\vec{x}(t),t)+\xi_j(t)+h_j(t),
\label{eq:general-Langevin}
\end{equation}  
where $\gamma_j$ is the friction coefficient for the variable $x_j$,  $F_j$  is a driving force that depends on the system configuration $\vec{x}=(x_1,x_2,...)$ and external driving,  $\xi(t)$ the zero-mean white Gaussian noise with variance $2\gamma_j T$.  The Boltzmann constant $k_B$ is set to be 1 throughout this article.   We assume that the external driving is such that the system reaches a NESS in the time scale much larger than the characteristic operation time.    $h_j(t)$ is a perturbative force that is applied only when we want to measure the linear response function of the system, defined to be $R_{\dot{x}_j}(t-\tau)\equiv \delta \langle \dot{x}_j(t)\rangle/\delta h_j(\tau)$.  Then,    the average heat dissipation rate  through the frictional motion of $x_j$ is given by $J_j\equiv \langle [\gamma_j\dot{x}_j(t)-\xi_j(t)]\circ \dot{x}_j(t)\rangle_{ss}$, where $\circ $ denotes the Stratonovich integral~\cite{sekimoto1998langevin} and $\langle \cdot\rangle_{ss}$ denotes the steady state ensemble.    In this general Langevin setting,  the Harada-Sasa equality states that~\cite{harada2005equality,harada2006energy}
\begin{equation}
J_j=\gamma_j  \left \{ \langle \dot{x}_j \rangle_{ss}^2+ \int_{-\infty}^\infty \frac{d\omega}{2\pi} [\tilde{C}_{\dot{x}_j}(\omega)-2T\tilde{R}_{\dot{x}_j}'(\omega)] \right \}.
\label{eq:HS}
\end{equation}
Here,  the prime denotes the real part,    $\tilde{R}_{\dot{x}_j}(\omega)$ is the Fourier transform of the response function $R_{\dot{x}_j}(t-\tau)$, and $\tilde{C}_{\dot{x}_j}(\omega)$ is the Fourier transform of the correlation function  $C_{\dot{x}_j}(t-\tau)\equiv \langle [\dot{x}_j(t)-\langle \dot{x}_j\rangle_{ss}][\dot{x}_j(\tau)-\langle \dot{x}_j\rangle_{ss}]\rangle_{ss}$.  The Fourier transform for a general function $g(t)$ is defined to be $\tilde{g}(\omega)\equiv \int_{-\infty}^\infty g(t)\exp(i\omega t)dt$,  with $i$ being the imaginary unit.    In the special case of equilibrium,   $\tilde{C}_{\dot{x}_j}(\omega)=2T\tilde{R}_{\dot{x}_j}'(\omega)$ due to FRR,  and the mean drift $\langle \dot{x}_j\rangle_{ss}$ and heat dissipation rate $J_j$ also vanishes.   In the steady state,  the total entropy production rate  $\sigma$ of the system is related to the heat dissipation rate through 
\begin{equation}
\sigma=\frac{1}{T}\sum_j J_j.
\end{equation}
This is the basis for estimating entropy production rate through analyzing the FRR violation for each channel $x_j$.

\section{Generalizing Harada-Sasa equality}
\label{sect:generalization_HS}

 So far,  the condition under which the generalized Harada-Sasa equality holds in Markov jumping systems  is still not clear.   Here,  we present a systematic derivation of the generalized Harada-Sasa equality Eq.~(\ref{eq:GHSE-2}) and Eq.~(\ref{eq:GHS}),  and identify their range of applicability. 

\subsection{General equality concerning FRR violation}

Consider a general Markov process.    The transition from state $n$ to state $m$  happens with  rate $w_n^m$.   We assume that if $w_n^m>0$, then the reverse rate $w_m^n>0$.  The self-transition is prohibited, i.e., $w_n^n=0$.   The probability  $P_n(t)$ at state $n$ and time $t$ evolves according to the following master equation
\begin{equation}
 \frac{d}{dt}P_n(t)=\sum_m M_{nm}P_m(t),
 \label{eq:govMatrix}
 \end{equation}
 where $M$ is assumed to be  an irreducible transition rate matrix determined by $M_{nm}=w_m^n-\delta_{nm}\sum_k w_n^k$.    We consider that an external perturbation  $h$ modifies the transition rate in the following way
\begin{equation}
\tilde{w}_m^n=w_m^n\exp\left[h\frac{\mathcal{Q}_n-\mathcal{Q}_m}{2T}\right],
\label{eq:perturbation}
\end{equation}
which is a generalization of the way Langevin system is perturbed~\cite{diezemann2005fluctuation,maes2009response_inter}.  Here,  $\mathcal{Q}_m$ is a conjugate variable to perturbation $h$.

The linear response of an arbitrary observable $A$ for this Markov system is defined to be $R_{A}(t-\tau)\equiv \delta \langle A(t)\rangle/\delta h(\tau)$.  In the last decade,  much efforts have been devoted to study the relation between linear response and fluctuation in non-equilibirum steady state.  By using path-integral approach,  Baiesi \emph{et al.} have derived the following relation~\cite{baiesi2009fluctuations}:
\begin{equation}
R_{A}(t-\tau)=-\frac{\beta}{2}\langle \bar{v}(\tau) A(t)\rangle_{ss}+\frac{\beta}{2} \langle \dot{Q}(\tau)  A(t)\rangle_{ss}.
\label{eq:response-NESS}
\end{equation}
Here,   along a stochastic trajectory $n_t$,  $Q(t)\equiv \mathcal{Q}_{n_t}$ and  $\dot{Q}(t)$ is the corresponding  instantaneous change rate of $Q(t)$.   We define
\begin{equation}
\bar{\nu}_n\equiv \sum_{n'} \omega_n^{n'}(\mathcal{Q}_{n'}-\mathcal{Q}_n),
\label{eq:nu}
\end{equation}
  which measures the average change rate of $Q(t)$ conditioned at the initial state $n$.  Then,  $\bar{v}(\tau)\equiv \bar{\nu}_{n_\tau}$.   For equilibrium systems,  
$\langle \dot{Q}(\tau) A(t)\rangle_{eq}=-\langle \bar{v}(\tau)A(t)\rangle_{eq}$ when $t>\tau$ and $\langle \dot{Q}(\tau) A(t)\rangle_{eq}=\langle \bar{v}(\tau) A(t)\rangle_{eq}$ when $t<\tau$.  These relations reduce Eq.~(\ref{eq:response-NESS})  to FRR in equilibrium.

Now, we focus on a specific application of  Eq.~(\ref{eq:response-NESS}) by choosing the observable $A(t)$ the same as $\dot{Q}(t)$ and  setting $t=\tau$ to consider the immediate response $ R_{\dot{Q}}(0)$.   After a little rearrangement,  we derive 
\begin{equation}
\langle \bar{v}(t)  \dot{Q}(t)\rangle_{ss}= \langle \dot{Q}\rangle_{ss}^2+\int_{-\infty}^\infty [\tilde{C}_{\dot{Q}}(\omega)-2k_BT \tilde{R}'_{\dot{Q}}(\omega)] \frac{d\omega}{2\pi},
\label{eq:GHSE-2}
 \end{equation}
 where the auto-correlation function $C_{\dot{Q}}(t-\tau)\equiv \langle (\dot{Q}(t)-\langle \dot{Q}\rangle_{ss})(\dot{Q}(\tau)-\langle \dot{Q}\rangle_{ss})\rangle_{ss}$.    Assuming that the system jumps from state $n_{\tau_j^-}$ to $n_{\tau_j^+}$ at the transition time $\tau_j$,    $\dot{Q}(t)= \sum_j \delta(t-\tau_j)[\mathcal{Q}_{n_{\tau_j^+}}-\mathcal{Q}_{n_{\tau_j^-}}]$,  which takes non-zero value only at the transition time $\tau_j$. However, the observable $\bar{v}(t)\equiv \bar{\nu}_{n_t}$ is not-well defined at the time of  transition.  This makes the evaluation of the correlation $\langle \bar{v}(t)  \dot{Q}(t)\rangle$ nontrivial.  Here, we define  
 \begin{equation}
 \bar{v}(t)\equiv \frac{1}{2}[\bar{\nu}_{n_{t^+}}+\bar{\nu}_{n_{t^-}}],
 \label{eq:bar_v}
 \end{equation}
  which takes the medium value at the transition.  To evaluate $\langle \bar{v}(t)  \dot{Q}(t)\rangle$,  we only need to consider the transition events.  For an ensemble of transition    from state $n$ to $m$,   $\bar{v}(t)$ gives   $[\bar{\nu}_n+\bar{\nu}_m]/2$,  while $\dot{Q}(t)$ gives  $P_nw_n^m(\mathcal{Q}_n-\mathcal{Q}_m)$, with $P_nw_n^m$ the average rate for such transition to occur.    Then,  summing over all possible transitions,   we derive
 \begin{equation}
\langle  \bar{v}(t) \dot{Q}(t) \rangle_{ss} = \frac{1}{4} \sum_{n,m} (\bar{\nu}_n+\bar{\nu}_m)[P_m\omega_m^n-P_n\omega_n^m](\mathcal{Q}_n-\mathcal{Q}_m) ,
\label{eq:evaluation}
\end{equation}
where we have symmetrized the result which gives rise to an additional  factor $1/2$.   Here,  $\langle  \bar{v}(t) \dot{Q}(t) \rangle_{ss}$ is proportional to the net flux $P_m\omega_m^n-P_n\omega_n^m$,  which vanishes at equilibrium,  in agreement with our expectation.   We claim original contribution for the derivation of Eq.~(\ref{eq:GHSE-2}) and Eq.~(\ref{eq:evaluation}).    Note that  while the rhs of Eq.~(\ref{eq:GHSE-2}) is very similar to that of the Harada-Sasa equality,  the lhs of Eq.~(\ref{eq:GHSE-2}) cannot in general be interpreted as heat dissipation for the degree of freedom $Q(t)$. 

\subsection{Connection FRR violation to heat dissipation rate}

Here,  we justify our medium value interpretation concerning Eq.~(\ref{eq:bar_v}) by refering to Langevin cases.  
Consider the observable $\dot{x}_j(t)$ for the Langevin equation~(\ref{eq:general-Langevin}). Here, for simplicity we assume that the force does not have explicit time-dependence,  and the NESS is achieved by breaking detailed balance in other ways.     The average change rate for $x_j$  at the position $\vec{x}$ is given by $\bar{\nu}_{\vec{x}}=F_j(\vec{x})/\gamma_j$.     For a transition from $\vec{x}$ to $\vec{x}'=\vec{x}+2\delta \vec{x}$,   our medium value interpretation implies that 
\begin{equation}
\bar{v}=\frac{F_j(\vec{x})+F_j(\vec{x}')}{2\gamma_j }= \frac{1}{\gamma_j}F_j\left(\frac{\vec{x}+\vec{x}'}{2}\right)+O(\delta \vec{x}^2).
\end{equation}
The second line is obtained by Talyor expansion.  For a trajectory with update time $\delta t$,  $\delta \vec{x}^2\sim \delta t$  and $\dot{x}_j\sim \delta t^{-1/2}$.   If we use the temporal average to approximate the ensemble average,  we have 
\begin{equation}
\langle \bar{v}(t) \dot{x}_j(t)\rangle_{ss}=\frac{1}{\gamma_j}\langle F_j(\vec{x}(t))\circ \dot{x}_j(t)\rangle_{ss}+O(\sqrt{\delta t}). 
\end{equation}
Therefore,  our medium value interpretation recovers the Stratonovich interpretation in the limit of $\delta t\to 0$,  and Eq.~(\ref{eq:GHSE-2}) is reduced to  the Harada-Sasa equality Eq.~(\ref{eq:HS}).  

\begin{figure}
\includegraphics[width=6cm]{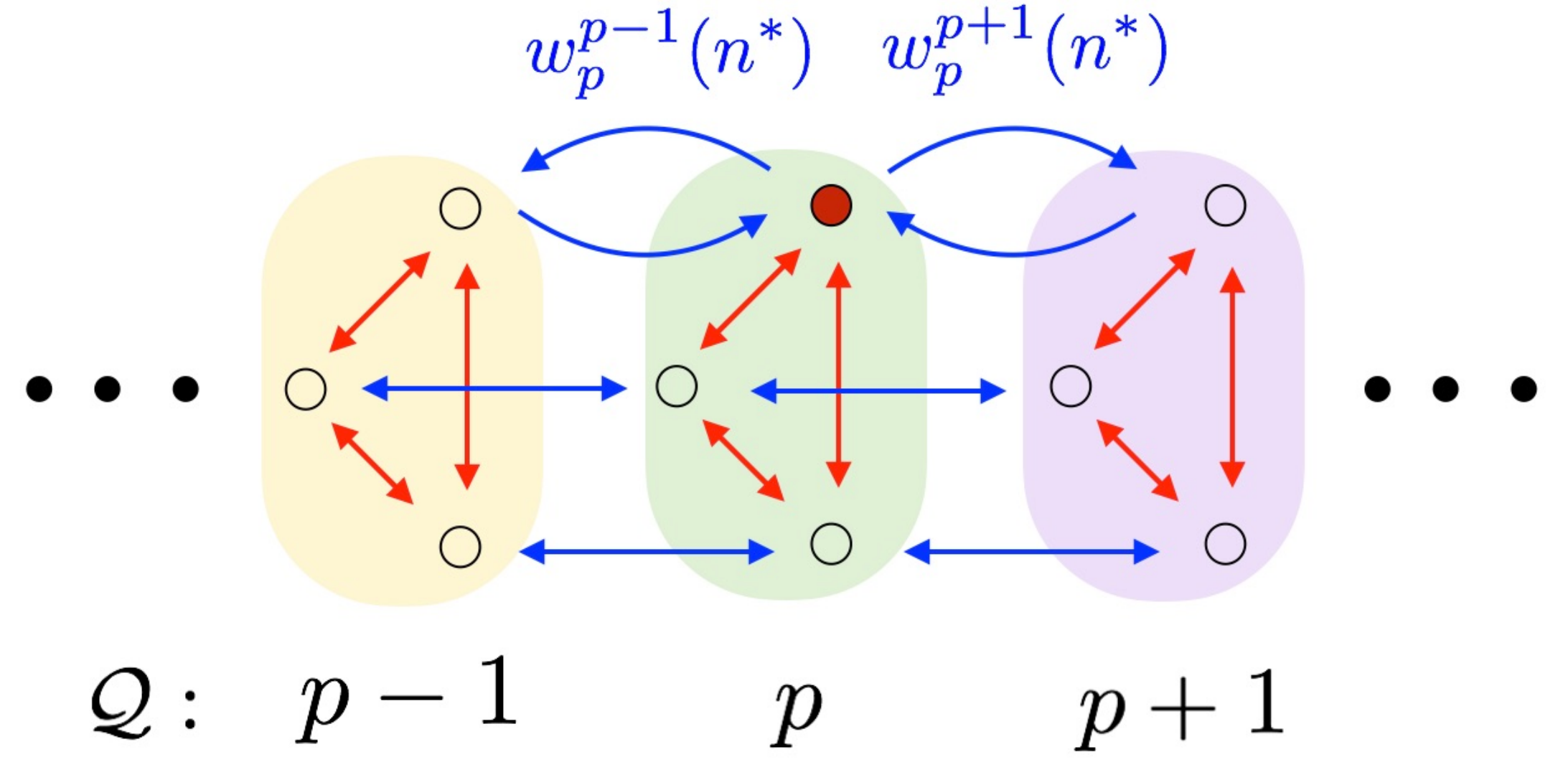}
\caption{Illustration of a multi-dimensional hopping process.   The black dots represents remaining blocks.  }
\label{fig:multi_hopping}
\end{figure}

 This result  can be easily generalized to a multi-dimensional hopping process.   The network should be easy to be decomposed into different directions.   Let us take FIG.~\ref{fig:multi_hopping} as an example, where the transition could be decomposed into the inter-block direction, indicated in blue arrows,  and intra-block direction, labeled in red.  The three-state block  represents a more complicated subnetwork that can also be decomposed into different directions. 
 
 Here,  we consider first how to capture the dissipation induced by the blue transitions.  The key is to choose a proper observable  $\dot{Q}(t)$ which  could just \emph{count} the blue transitions with a proper weight $\bar{v}(t)$ that relates to the dissipation during the corresponding transition.  To do so,  we set the conjugate variable  $\mathcal{Q}$ to be uniform  within each block  and make it change value by 1 when jumping to a neighboring block.      Let $n^*$ represents a label for states inside the block.    For the state $(p, n^*)$, which is  labeled in red in FIG.~\ref{fig:multi_hopping},   the average change rate for $\mathcal{Q}$ becomes
 \begin{equation}
 \bar{\nu}_{p,n^*}=w_p^{p+1}(n^*)-w_p^{p-1}(n^*)
 \end{equation}
  according to Eq.~(\ref{eq:nu}),  which gives the inherent property of this state.  Similar to the above discussion of Langevin processes,  $\bar{\nu}_{p,n^*}$ can be related to some kind of force, or dissipation per jump.  We denote  
  \begin{equation}
  \Delta S(n^*,p)\equiv \ln w_p^{p+1}(n^*)/w_{p+1}^{p}(n^*)
  \label{eq:Delta-S}
  \end{equation}
  as the entropy produced in the medium during the transition from state $(n^*,p)$ to $(n^*,p+1)$.  Then $T\Delta S(n^*,p)$ gives the heat dissipation for this jump.    Now,  if we assume that for all the blue transitions in FIG.~\ref{fig:multi_hopping} the transition rate satisfies the form 
\begin{subequations}\label{eq:interblock-rates}
\begin{eqnarray}
w_p^{p+1}(n^*)&=&\frac{1}{\tau_Q}\exp\Big((1-\theta) \Delta S(n^*,p) \Big)\\
 w_{p+1}^{p}(n^*)&=&\frac{1}{\tau_Q}\exp\Big( - \theta \Delta S(n^*,p)\Big),
\end{eqnarray}
\end{subequations} 
 where $\tau_Q$ is a constant number and $\theta$ is a load sharing factor.  Then,   
\begin{equation}
 \bar{\nu}_{p,n^*}=\frac{1}{\tau_Q}\left(\Delta S(p)+\theta [\Delta S(p-1)-\Delta S(p)]+O(\Delta S^2)\right),
\end{equation}
where we have suppressed for $\Delta S$ the dependence on  state $(n^*,p)$ for simplicity.  Assuming that  $|\Delta S|\ll 1$ and that $\Delta S$ varies slowly along $p$, i.e., $[\Delta S(p-1)-\Delta S(p)]\ll 1$,  we now successfully connect the average change rate of $\mathcal{Q}$ with the dissipation per jump: 
\begin{equation}
\bar{\nu}_{p,n^*}\approx  \frac{1}{\tau_Q} \Delta S(p,n^*).
\end{equation}
Combined with Eq.~(\ref{eq:evaluation}) and Eq.~(\ref{eq:Delta-S}),  the integral of the FRR violation for this observable is given by
 \begin{equation}
 \langle \bar{v}(t)  \dot{Q}(t)\rangle_{ss} \approx \frac{1}{\tau_Q}\sum_{n^*,p}(P_{p}^{ss}w_p^{p+1}-P_{p+1}^{ss}w_{p+1}^{p})\ln \frac{w_p^{p+1}}{w_{p+1}^{p}}.
  \label{eq:GHS}
  \end{equation}  
Therefore,   combined with Eq.~(\ref{eq:GHSE-2}), this equation implies that  the FRR violation of the observable $\dot{Q}$ captures the dissipation rate due to the inter-block transitions, as indicated by the blue arrows.   This equation along with Eq.~(\ref{eq:GHSE-2})  constitute our generalized Harada-Sasa equality.  Here,  $T\tau_Q$ is the corresponding friction coefficient $\gamma_Q$.

Now,  we make some comments related to Eq.~(\ref{eq:GHS}). 

(a) Key assumptions include that   all the inter-block transitions share the same timescale $\tau_Q$ specified by Eq.~(\ref{eq:interblock-rates}),  that the dissipation  per jump is relatively small compared with the thermal energy $T$,  and that the dissipation changes slowly for neighboring jumps between blocks.  However,  the load sharing factor $\theta$ is not required to be 1/2,  which was assumed previously  by Lippiello \emph{et al.}~\cite{Lippiello2014fluctuation}. 

 (b)  The observable $Q$ is chosen to be a linear function along the block hopping direction such that $\dot{Q}(t)$ would count the transitions.  Even if the actual observable in the experiment is not such a linear form given in FIG.~\ref{fig:multi_hopping},  we can map the observed trajectory $\dot{Q}_{old}$ to a new one $\dot{Q}_{new}$ associated with a properly designed  observable  at the stage of data analysis. 
 
    (c) To access the dissipation rates due to the inter-block transitions,  we have made no assumptions about the transitions inside the blocks.  However, in order to access the total dissipation rates,  we need to devise other observables to count the transitions inside the blocks and these transitions should satisfiy similar constraints. 
    
     (d)  In certain cases,  although $\Delta S$ is not always small,  the probability flux becomes dominant  only around the transitions where  $\Delta S$ is small.   Therefore,   Eq.~(\ref{eq:GHS}) may also be valid,  as illustrated later by our example in FIG.~\ref{fig:HS-adaptation-2}.

\section{Correlation and response in a general Markov system}
\label{sect:CR}
\subsection{Setup}
Here,  assumping general Markov processes,  we derive the velocity correlation spectrum Eq.~(\ref{eq:Cvelo_fre}) and response spectrum Eq.~(\ref{eq:Cvelo_fre}) for a general observable,  and its FRR violation spectrum Eq.~(\ref{eq:velo-FRR-vio}).  Our strategy is to project these spectrum in the eigenspace of the evolution operator.

 Consider a general Markov process the similar as introduced above,  except that it has only finite states, say $N$ states.    The $j$-th left and right eigenmodes,  denoted as $x_j(n)$ and $y_j(n)$ respectively,  satisfy the eigenvalue equation 
 \begin{subequations} \label{eq:xy}
 \begin{eqnarray}
 \sum_m M_{nm}x_j(m)&=-\lambda_j x_j(n)\\
  \sum_m y_j(m)M_{mn}&=-\lambda_j y_j(n), 
 \end{eqnarray}
 \end{subequations}
 where the minus sign is introduced to have a positive ``eigenvalue" $\lambda_j$~\cite{van1992stochastic}.  These eigenvalues are arranged in the ascending order by their real part, i.e., $\text{Re}(\lambda_1)\le \text{Re}(\lambda_2)\le \cdots$.     This system will reach a unique stationary state associated with  $\lambda_1=0$, where $y_1=1$ and  $x_1(m)=P_{m}^{ss}$ due to probability conservation.  With the proper normalization $\sum_m P_m^{ss}=1$,  the eigenmodes satisfy the  orthogonal relations $\sum_m x_j(m)y_{j'}(m)=\delta_{jj'}$ and completeness relations $\sum_j x_j(n)y_j(m)=\delta_{nm}$.

 \subsection{Correlation spectrum}
Consider first  the auto correlation function $C_Q(t-\tau)$ for the observable $Q(t)\equiv \mathcal{Q}_{n_t}$,  which can be reformulated  in the following form 
\[
C_{Q}(t-\tau)= \sum_{n,n'}  \mathcal{Q}_{n}\mathcal{Q}_{n'} P(t-\tau;n,n')P_{n'}^{ss} -\langle Q\rangle_{ss}^2,
\]
where $P(t;n,n')$ is the probability that a system starting in state $n'$ at time $t=\tau$ would reach state $n$ at time $t$.  An expansion in the eigen space gives  $P(t;n,n')=\sum_{j} y_j(n') e^{-\lambda_j |t-\tau|} x_j(n)$,  which satisfies the master equation Eq.~(\ref{eq:govMatrix}) and the initial condition $P(0;n,n')=\delta_{nn'}$.   Introducing the weighted average of $\mathcal{Q}$ in the $j$-th eigenmodes,   i.e.,  $\alpha_j\equiv\sum_n \mathcal{Q}_nx_j(n)$ and $\beta_j \equiv \sum_{n}\mathcal{Q}_{n}  y_j(n)P_n^{ss}$,   the correlation function is expanded in the eigenspace, i.e., 
\begin{equation}
C_{Q}(t-\tau) =\sum_{j=2}^N \alpha_j \beta_j e^{-\lambda_j |t-\tau|}.
\label{eq:Corr1}
\end{equation}  
   The contribution of the ground state ($j=1$) cancels that of the mean deviation, i.e., $\langle Q\rangle^2_{ss}$.  Note that the stationarity leads to     $C_Q(t-\tau)=C_Q(\tau-t)$,  which constrains Eq.~(\ref{eq:Corr1}) through the absolute term $|t-\tau|$.  This can be derived by 
The correlation function $C_{\dot{Q}}(t-\tau)$ for the velocity observable $\dot{Q}(t)$ can be obtained by the transformation
\[
    C_{\dot{Q}}(t-\tau)=\frac{\partial^2 C_{Q}(t-\tau)}{\partial \tau\partial t}.
\]
In the Fourier space,  we have 
\begin{equation}
  \tilde{C}_{\dot{Q}}(\omega)=\sum_{j=2}^N 2\alpha_j\beta_j\lambda_j \Big[1-\frac{1}{1+(\omega/\lambda_j)^2}\Big] ,
 \label{eq:Cvelo_fre}
  \end{equation}
  which is generally valid for a system in NESS. 

\subsection{Response spectrum}
The response spectrum can be obtained by studying the response of the system to   a  periodic perturbation.  Consider  $h=h_0 \exp(i \omega t  )$ with $h_0$ a small amplitude and $ i$ the imaginary unit.  Up to  the first order of $h_0$, 
 the  transition rate matrix $\tilde{M}$ is modified as
\[\tilde{M}=M+h_0 \partial_h \tilde{M} \exp(i\omega t  )+O(h_0^2).\]
After a sufficiently long time,  the system reaches a distribution with a periodic temporal component that has time-independent amplitude:
\[\tilde{P}_m=P_m^{ss}+h_0 P^{(1)}_m\exp(i\omega t )+O(h_0^2).\]
Here,  with the stationary condition of the zeroth term, i.e., $\sum_m M_{nm}P_m^{ss}=0$,  the new master equation $d\tilde{P}_m/dt=\sum_n \tilde{M}_{mn}\tilde{P}_n$ determines the first order correction of the distribution
 \[P^{(1)}=-\frac{1}{M-i\omega  } \partial_h M P^{ss},\]
  written in a Matrix form.  By  introducing     
  \begin{equation}
  B_n\equiv\sum_{m}\partial_h \tilde{M}_{nm} P^{ss}_m,
  \label{eq:general-Bn}
  \end{equation}
  and the weighted average of $B$ in the $j-th$ eigenmode,  i.e.,  $\phi_j \equiv \sum_{n} B_{n} y_j(n)$,   the linear order variation is expressed as
\[
P^{(1)}_n=\sum_{j=2}^N \frac{1}{\lambda_j+i\omega} \phi_j x_j(n).
\]
The first mode disappears as one can check that $\phi_1=\sum_mB_my_1(m)=\sum_m B_m=0$, because the ground state does not contain dynamic information.  
Finally,  for a state-dependent observable $Q(t)\equiv\mathcal{Q}_{n_t}$,   its response spectrum is given by 
\begin{equation}
\tilde{R}_Q(\omega)=\sum_n \mathcal{Q}_nP^{(1)}_n=\sum_{j=2}^N \frac{\alpha_j\phi_j}{\lambda_j+i\omega}.
\label{eq:Rq-general-fre}
\end{equation}
By using the transformation $ R_{\dot{Q}}(t)=dR_{Q}/dt$ or $\tilde{R}_{\dot{Q}}(\omega)=i\omega \tilde{R}_{Q}(\omega)$,   we obtain the desired response spectrum $\tilde{R}_{\dot{Q}}(\omega)$ for the velocity observable $\dot{Q}(t)$, i.e., 
  \begin{equation}
   \tilde{R}_{\dot{Q}}(\omega)=\sum_{j=2}^N \alpha_j\phi_j\Big[1- \frac{1- i(\omega/\lambda_j) }{1+(\omega/\lambda_j)^2}\Big].
   \label{eq:Rvelo_fre}
   \end{equation}
  For the perturbation form in Eq.~(\ref{eq:perturbation}),  we have
   \begin{equation}
B_n = \sum_m[ w_m^nP_m^{ss}+w_n^mP_n^{ss} ](\mathcal{Q}_n-\mathcal{Q}_m)/2T,
\label{eq:Bn-main}
\end{equation}
which gives the flux fluctuation of the conjugate variable $\mathcal{Q}_n$ at state $n$. 

\subsection{Useful relations}
We find that the coefficients always satisfy the following relation
\begin{equation}\label{eq:sumrule-general}
\sum_{j=2}^N \alpha_j (\lambda_j \beta_j-T\phi_j)=0,
\end{equation}
which is called the \emph{sum rule}.  See Appendix~\ref{app:sum-rule} for the derivation.   Combining Eq.~(\ref{eq:Rvelo_fre}) and Eq.~(\ref{eq:Cvelo_fre}),  this relation leads to FRR in high frequency domain ($\omega\gg\lambda_N$), i.e., 
\begin{equation}
 \tilde{C}_{\dot{Q}}(\omega)=  2T\tilde{R}'_{\dot{Q}}(\omega).
\end{equation}
This is consistent with our intuition that when the frequency is much higher than the characteristic rate of the system the correlation and response spectrum of the system  only reflects the property of the thermal bath that is in equilibrium.

Combining Eq.~(\ref{eq:Cvelo_fre}) and Eq.~(\ref{eq:Rvelo_fre}),   the FRR violation spectrum for a velocity observable $\dot{Q}$ can be generally written as 
  \begin{equation}
 \tilde{C}_{\dot{Q}}(\omega)-2T\tilde{R}'_{\dot{Q}}(\omega)=2\sum_{j=2}^N \alpha_j \frac{T\phi_j-\beta_j\lambda_j}{1+(\omega/\lambda_j)^2}.
 \label{eq:velo-FRR-vio}
 \end{equation}
Although the involved coefficients and eigenvalues in the rhs are probably complex numbers,  the summation over all the eigenmodes guarantees a real violation spectrum,  as shown in Appendix~\ref{app:realPart}.   In the limit $\omega\to 0$,  FRR is also valid since both the correlation and response  becomes zero,  as evident from Eq.~(\ref{eq:Cvelo_fre}) and Eq.~(\ref{eq:Rvelo_fre}).    The integral of the FRR violation,  denoted as $\Delta_Q$,  is given by 
\begin{eqnarray}
\Delta_Q&=&\int_{-\infty}^\infty  \left (\tilde{C}_{\dot{Q}}(\omega)-2T\tilde{R}'_{\dot{Q}}(\omega)\right) \frac{d\omega}{2\pi}\nonumber\\
&=&\sum_j \lambda_j\alpha_j \left(T\phi_j-\beta_j\lambda_j\right).
\label{eq:Delta-Q-expansion}
\end{eqnarray}
The Harada-Sasa equality suggests that this quantity is  related to the dissipation through the motion of $Q$.

We also find that   the detailed balance  $w_n^mP_n^{eq}=w_m^nP_m^{eq}$ is  equivalent to 
\begin{equation} \label{eq:detailed-eigenmode}
\lambda_j\beta_j^{eq}=T\phi_j^{eq},
\end{equation}
which ensures that FRR be satisfied in all frequency domain.    This relation is a general result independent of the perturbation form proposed in Eq.~(\ref{eq:perturbation}),  as proved in  Appendix~\ref{app:FRReigenmode}.  Therefore,  we may also talk about detailed balance in the eigenspace,  and an eigenmode contributes to dissipation only when it violates the detailed balance,  according to Eq.~(\ref{eq:Delta-Q-expansion}).

\section{Markov processes with time scale separation }
\label{sect:time scale}
Now,  we consider Markov processes with time scale separation.     We assume that the state space can be grouped into $K$ different coarse-grained subspaces,   denoted as $p$ or $q$.  A microscopic state $k$ ($l$) within coarse-grained $p$ ($q$) is denoted  as $p_k$ ($q_l$),  which serves as an alternative to our previous state notation $n$ or $m$.  We assume fast relaxation ($\sim \tau_f$) within the subspace of a coarse-grained state and slow `hopping ($\sim\tau_s $)   to other subspaces associated with a different  coarse-grained state.  The competition of these two processes defines a dimensionless parameter $\epsilon\equiv\tau_f/\tau_s$.  A typical example of this Markov system is illustrated in FIG.~\ref{fig:general-model}.     Such an assumption implies that  the transition rate matrix of the system can be decomposed as\begin{equation}
M_{p_kq_l}=\frac{1}{\epsilon}\delta_{pq}M^{p}_{kl}+M^{(0)}_{p_kq_l},
\label{eq:master-M}
\end{equation}
where  $M^{p}_{kl}$ ($\sim 1$)  describes  a rescaled  ``internal"  Markov process within  the  same coarse-grained state $p$, and $M^{(0)}_{p_kq_l}$  for jumps connecting different coarse-grained states.  The transition rate  from state $k$ to $l$ for  $M^p$ is  denoted as  $w_k^l(p)$ ($\sim 1$). 

   The goal of this section is to obtain Eq.~(\ref{eq:velo-FRR-vio-fast}),  concerning  the structure of the FRR violation spectrum for a general  observable in this system,  and also Eq.~(\ref{eq:velo-FRR-vio-slow})  for an observable that could only resolve different coarse-grained state,  which is of particular interest.       Below,   we start by  obtaining the eigenmodes of $M$ by perturbation theory.

\subsection{Perturbation analysis for eigenmodes}

%%%%%%%%%%%%%
\begin{figure}
\centering
\includegraphics[width=8cm]{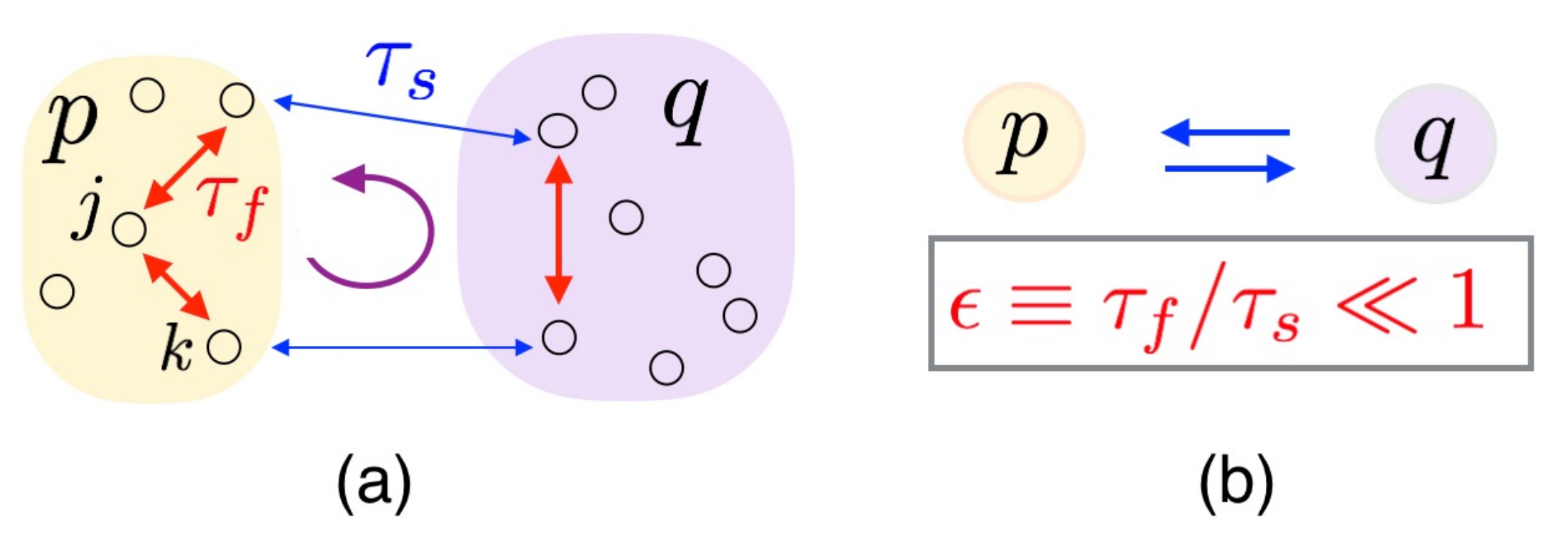}
\caption{(a) Our system with two time scales.  One of the closed cycles that breaks time-reversal symmetry is indicated, which is responsible for hidden entropy production.     (b) The corresponding effective dynamics. }
\label{fig:general-model}
\end{figure}
%%%%%%%%%%%%%

 We  write the eigenmodes and the corresponding eigenvalue as Taylor series  in $\epsilon$, i.e.,
\begin{eqnarray*}
x_j&=&x_j^{(0)}+\epsilon x_j^{(1)}+O(\epsilon^2)\\
y_j&=&y_j^{(0)}+\epsilon y_j^{(1)}+O(\epsilon^2)\\
\lambda_j&=&\epsilon^{-1} \lambda_j^{(-1)}+\lambda_j^{(0)}+O(\epsilon^2),
\end{eqnarray*} 
and    substitute these expansions into the  eigenvalue equations Eq.~(\ref{eq:xy}).    The leading order equations in $\epsilon $  are given by  
\begin{subequations}\label{eq:zero-xy}
\begin{eqnarray}
 \sum_{l} M^{p}_{kl}x^{(0)}_j(p_l)&=-\lambda_j^{(-1)}x_j^{(0)}(p_k)\\
  \sum_{l}y^{(0)}_j(p_l) M^{p}_{lk}&=-\lambda_j^{(-1)}y_j^{(0)}(p_k).
\end{eqnarray}
\end{subequations}
Therefore,  the eigenmodes of the intra-block transition rate matrix $\sum_p M^p$ are the same as the leading order eigenmodes of $M$.   At $\lambda_j^{(-1)}\neq 0$,  these eigenmodes are in general non-degenerate and decay quickly at the time scale $\tau_f$,  thus called the fast modes. 

 The remaining $K$ slow modes corresponding to $\lambda_j^{(-1)}=0$ are degenerate now,  and the lift of this degeneracy is due to the inter-block transition,  which requires the next order perturbation analysis, i.e., 
\begin{eqnarray*}
  \sum_l M^{p}_{kl}x^{(1)}_j(p_l)+   \sum_{q_l} M^{(0)}_{p_kq_l}x^{(0)}_j(q_l)&=-\lambda_j^{(0)}x_j^{(0)}(p_k)\\
 \sum_l y^{(1)}_j(p_l)M^{p}_{lk}+ \sum_{q_l}y^{(0)}_j(q_l)M^{(0)}_{q_lp_k}&= -\lambda_j^{(0)}y_j^{(0)}(p_k).
\end{eqnarray*}
  Although the lhs (left hand side) depends on the unknown first order correction of the eigenmodes,  we can eliminate these unknown terms by projecting the first equation on the left stationary mode of $M^p$, i.e., denoted as $y_1^p=1$,  and projecting the second equation on the right stationary mode of $M^p$,  denoted as  $P(k|p)$ which satisfies  the normalization $\sum_k P(k|p)=1$ and    
\begin{equation}
\sum_k M^p_{lk} P(k|p)=0.
\end{equation}
We also introduce the following ansatz  for  these slow modes 
\begin{equation}
x^{(0)}_j(p_k)=\wh{x}_j(p)P(k|p),\quad y^{(0)}_j(p_k)=\wh{y}_j(p),
\label{eq:zero-th-xy}
\end{equation}
which simply means that the eigenmodes are stationary under intra-block transition but have a modulation at the inter-block level.  Then,  we obtain   a reduced eigenvalue equations for these modulation amplitudes, i.e.,  
\begin{subequations}\label{eq:coarse-grain-xy}
\begin{eqnarray}
\sum_{q} \wh{M}_{pq}\wh{x}_j(q)&=- \lambda_j^{(0)} \wh{x}_j(p) \\
 \sum_{q}\wh{y}_j(q)\wh{M}_{qp}&=-\lambda_j^{(0)}\wh{y}_j(p).
\end{eqnarray}
\end{subequations}
Here,  the emergent transition rate matrix on the coarse-grained state space is given by 
\begin{equation}
\wh{M}_{pq}\equiv\sum_{k,l} M_{p_kq_l}^{(0)}P(l|q),
\label{eq:effective-M}
\end{equation}
which is exactly due to  a projection by the left and right stationary modes of the intra-block transition rate matrix.   The projection procedure is effectively a coarse-graining over the microscopic states within the same block.    Therefore,   the effective matrix $\wh{M}$ removes the $K$-fold degeneracy of the slow modes,  and determines its leading order behavior.

Now we summarize the  non-degenerate leading order term of the eigenmodes  of $M$.   These eigenmodes are split into two classes in terms of its relaxation time scale:  fast modes that relax at the time scale $\sim \tau_f$ and slow ones at the time scale $\sim \tau_s$.  This is illustrated in FIG.~\ref{fig:general-model}(c) for an illustrative example.   For a fast mode,  its leading order term is localized  within a certain coarse-grained state $p$.  We may  denote the non-stationary eigenmodes  of  $M^p$ as $x_j^p$ and $y_j^p$,  and the corresponding eigenvalue as $\lambda_j^p$.  They  satisfy the following orthogonal relations 
  \begin{equation}
  \sum_k x_j^q(k)=0,\qquad  \sum_k y^q_j(k) P(k|q)=0.
  \label{eq:orthogo}
  \end{equation}
The first relation can be understood from probability conservation,  while the second one due to stationarity of the ground state.   Then,   this fast mode  can be expressed as   $(j>K)$
      \begin{subequations}\label{eq:fast-mode-corresp}
  \begin{eqnarray}
\lambda_j &=&\epsilon^{-1}\lambda_j^p+O(1)\\
x_j(q_k)&=&\delta_{pq}x_j^p(k)+ O(\epsilon)\\
 y_j(q_k)&=&\delta_{pq}y_j^p(k)+  O(\epsilon).
\end{eqnarray}
  \end{subequations} 
To express the slow modes,  we introduce the eigenmodes of   $\wh{M}$  as $\wh{x}_{j}$ and $\wh{y}_{j}$,  with corresponding  eigenvalue $\wh{\lambda}_{j}$.  Then,    the slow modes become $(j\le K)$  
\begin{subequations}\label{eq:slow-mode-corresp}
\begin{eqnarray}
 \lambda_j&=&\wh{\lambda}_j +O(\epsilon)\\
x_j(p_k)&=&\wh{x}_j(p)P(k|p)+O(\epsilon)\nonumber\\
\\
\quad  y_j(p_k)&=&\wh{y}_j(p)+O(\epsilon).
\end{eqnarray}
\end{subequations}
In particular, the stationary distribution of the original system becomes   
\begin{equation}
P^{ss}_{p_k}=\wh{P}^{ss}_pP(k|p)+O(\epsilon),
\end{equation}
  with  $\wh{P}^{ss}_p$ the normalized  stationary distribution for the coarse-grained Markov system. 

The leading order results are sufficient for the following discussion.  The first order correction is discussed in Appendix~\ref{sect:higher-correction}.      We note that the first order correction is in general very complicated.  The correction of the fast modes involves only 1-step transition to all the other eigenmodes,  while for the slow modes  the correction also involves 2-step transition,  i.e. first to a fast mode and then back to a different slow mode.  The latter is generic in  degenerate perturbation~\cite{sakurai1985modern}. 

%  which is essentially the same as the time-independent perturbation in quantum mechanics~\cite{sakurai2014modern}. 

%%%%
\subsection{FRR violation spectrum for fast observables}
\label{sect:FRRvio}
For a general conjugate variable $\mathcal{Q}^f_{p_k}$ that depends on microscopic states,  its corresponding velocity observable $\dot{Q}^f$ moves at a fast time scale $\sim \tau_f$,  thus classified as a fast variable and emphasized by the superscript $f$.   To obtain the property of  its violation spectrum,  we  study the  asymptotic behavior of the corresponding projection coefficients $\alpha_j^f$, $\beta_j^f$ and $\phi_j^f$ with the help of  the eigenmodes obtained in the previous section.

\begin{figure}
\centering
\includegraphics[width=6cm]{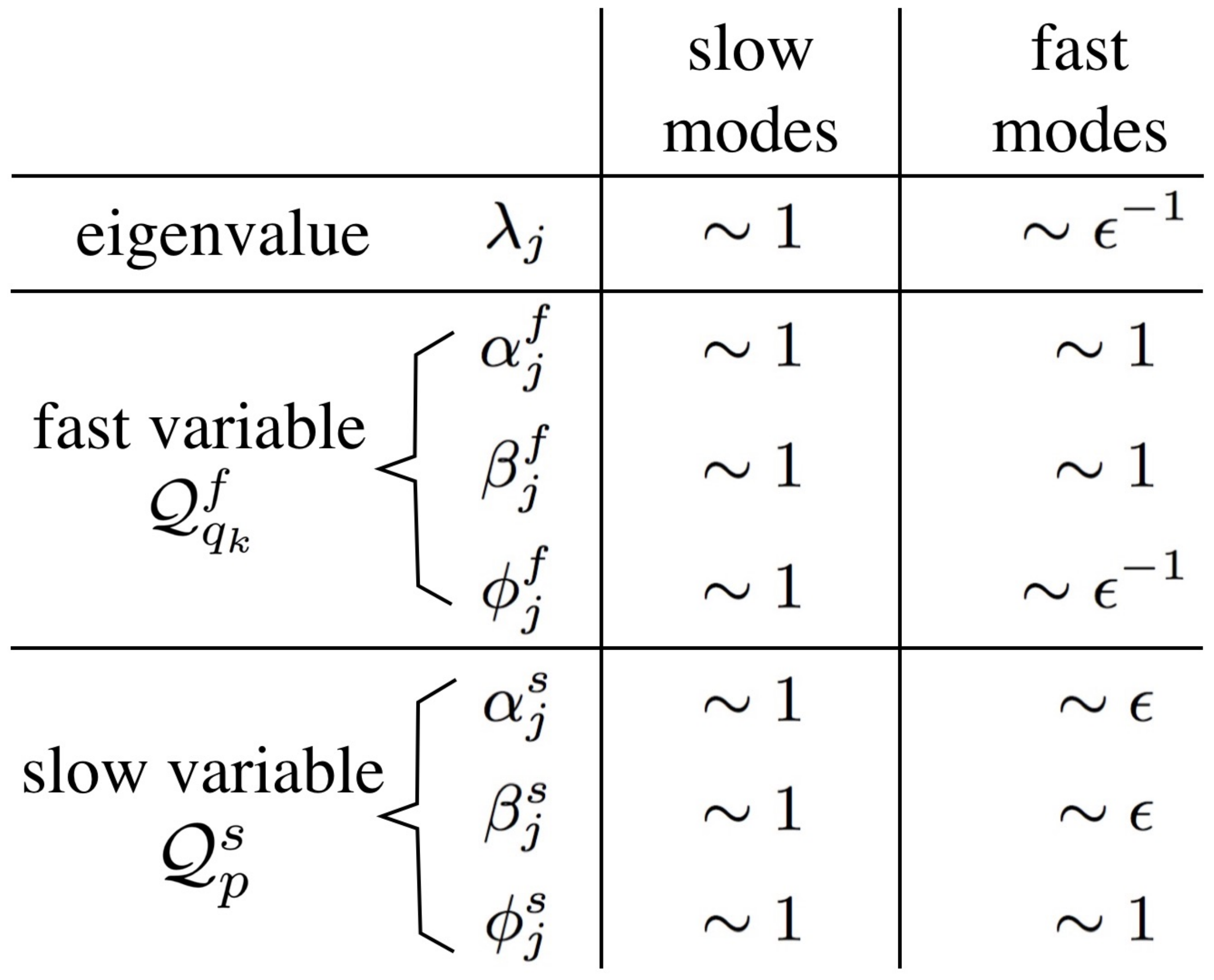}
\caption{Overview of asymptotic behavior of the crucial parameters of correlation and response spectrum in this Markov system with time scale separation.   }
\label{fig:asymp}
\end{figure}

We first estimate their magnitude in the time scale separation limit $\epsilon\to 0$.   Since the leading order terms for  $\alpha_j^f$ and $\beta_j^f$ do not vanish for such a fast observable,    we roughly obtain their magnitude as 
\begin{equation}
\alpha_j^f\sim 1 ,\quad \beta_j^f\sim 1.
\end{equation}
More delicate results up to first order correction can be obtained immediately by using Eq.~(\ref{eq:fast-mode-corresp}) and Eq.~(\ref{eq:slow-mode-corresp}).  To obtain the magnitude of    $\phi_j^f\equiv\sum_{p_k} y_j(p_k) B_{p_k}$,  we need to understand $B_{p_k}$ first.  It  can be expanded as  [Eq.~(\ref{eq:Bn-main})]
\begin{equation}
B_{p_k}=\epsilon^{-1}B^p_{k}+B_{p_k}^{(1)},
\label{eq:B-split}
\end{equation}
where the leading order term  $\epsilon^{-1}B^p_{k}\sim \epsilon^{-1}$ describes fast flux fluctuation  within coarse-grained state $p$,  as determined by $M^p$;  at the same time $B_{p_k}^{(1)}\sim 1$ gives the slow flux fluctuation  due to inter-block fluctuation, as determined  by $M^{(0)}$.   The projection of $B_{p_k}$ on  fast modes generally have a large value, i.e., 
 \begin{equation}
  \phi_j^f\sim \epsilon^{-1},\quad j> K.
 \end{equation}
 However,  its projection on slow modes  are greatly supressed,  
  \begin{equation}
 \phi_j^f\sim 1,\quad j\le K,
 \end{equation}
because of  the quasi uniformness of the slow modes in a given coarse-grained state, i.e.,  $y_j(p_k)\approx \wh{y}(p)$, and  the  conservation   of flux fluctuation within each coarse-grained state, i.e., $\sum_k B^p_k=0$. The magnitude of these  projection coefficients are listed in FIG.~\ref{fig:asymp}.

Since each fast mode  has a counterpart from a certain internal transition rate matrix $M^p$,  their  projection coefficients also share this connection.   For the Markov process described by $M^p$,  we may also introduce the  projection coefficients  $\alpha_j^q\equiv\sum_k x^q_j(k)\mathcal{Q}_{q_k}^f$, $\beta_j^q\equiv\sum_k y^q_j(k)\mathcal{Q}_{q_k}^f P(k|q)$ and $\phi_j^q\equiv \sum_k y^q_j(k)B(k|q)$.  Here,  $B(k|q)$ is the flux fluctuatioin for this subsystem,  which is found to satisfy the relation $B^q_k=\wh{P}^{ss}_pB(k|q)+O(\epsilon)$.   We find that these coefficients are connected to those of the fast mode by   ($j>K$) 
\begin{subequations} \label{eq:eff-org-fast-2}
\begin{eqnarray}
\alpha_j^f&=&\alpha^q_j+O(\epsilon)\\
\beta_j^f&=&\wh{P}^{ss}_q\beta^q_j+O(\epsilon)\\
\phi_j^f&=&\epsilon^{-1} \wh{P}_p^{ss}\phi^q_j+O(1).
\end{eqnarray}
\end{subequations}
The FRR violation spectrum  for this fast velocity observable $\dot{Q}^f$ can be split into the contribution of each fast internal Markov processes [Eq.~(\ref{eq:velo-FRR-vio})] and a correction due to the slow transition across different coarse-grained state, i.e.,  
  \begin{eqnarray}
 \tilde{C}_{\dot{Q}^f}-2T\tilde{R}'_{\dot{Q}^f}&=&\frac{2}{\epsilon}\sum_{j>K} \wh{P}^{ss}_p \left[\alpha_j^q \frac{T\phi_j^q-\beta_j^q\lambda_j^q}{1+(\epsilon\omega/ \lambda_j^q)^2}\right]+V_f(\omega),\nonumber\\
 \label{eq:velo-FRR-vio-fast}
 \end{eqnarray}
 where the summation is  first over all the non-stationary modes of $M^p$,   and then over all the coarse-grained state $p$.     The correction term is given by 
 \begin{equation*}
 V_f(\omega)=2\sum_{j\ge 2} \alpha_j^f\frac{T\phi_j^f-\beta_j^f\lambda_j}{1+[\omega/\lambda_j]^2}-\frac{2}{\epsilon}\sum_{j>K} \wh{P}^{ss}_p \left[\alpha_j^q \frac{T\phi_j^q-\beta_j^q\lambda_j^q}{1+(\epsilon\omega/ \lambda_j^q)^2}\right].
 \end{equation*}
The two terms of the rhs have the same divergence of order $\epsilon^{-1}$,  which cancels each other due to the mapping relation Eq.~(\ref{eq:eff-org-fast-2}).  Therefore, $ V_f(\omega)$ is of order 1 and is well-defined in the time scale separation limit $\epsilon\to 0$.   
  Generally,  $V_f$ vanishes in the high frequency region $\omega\gg\lambda_N$,  as expected from the sum rule.   It also vanishes in the low frequency region for our setup with a finite state space.    In the intermediate frequency region $\tau_s^{-1}\ll \omega\ll \tau_f^{-1}$, the contribute from slow modes is negligible and 
 \begin{equation}
 V_f\simeq 2\sum_{j>K}  \alpha_j^f \left[T\phi_j^f-\beta_j^f\lambda_j^f\right]\xrightarrow{\epsilon\to 0} \text{const}.
 \end{equation}
 The finite limit is reached due to the mapping relations Eq.~(\ref{eq:eff-org-fast-2}).   
 
The FRR violation integral $\Delta_{Q^f}$ also splits into two terms 
\begin{equation}
\Delta_{Q^f}=\frac{1}{\epsilon} \sum_q \wh{P}^{ss}_q \Delta_{Q^f}^q+ \Delta_{Q^f}^{(1)},
\end{equation}
where $\Delta_{Q^f}^q$ is contributed by the FRR violation integral within mesoscopic state $q$, associated with $M^q$,  and $\Delta_{Q^f}^{(1)}$ is the correction term contributed by integrating over $V_f(\omega)$.   Both $\Delta_{Q^f}^q$ and $\Delta_{Q^f}^{(1)}$ scale as $\epsilon^{-1}$ since the violation plateau spans up to frequency $1/\tau_f\sim \epsilon^{-1}$.   
Therefore,  the leading order term $\frac{1}{\epsilon} \sum_q \wh{P}^{ss}_q \Delta_{Q^f}^q\sim \epsilon^{-2}$ and $ \Delta_{Q^f}^{(1)}$ is a negligible  correction.  Indeed,  this leading order term implies a diverging dissipation rate of the system,  which is not quite realistic.  It is then natural to assume that detailed balance is satisfied within each coarse-grained state,   i.e., $\phi_j^q=\beta_j^q\lambda_j^q$,  or equivalently $\Delta_{Q^f}^q=0$.   The FRR violation spectrum of this fast observable is then the same as $V_f(\omega)$,  which is illustrated in FIG.~\ref{fig:illustration_FRRviolation}(a).  

%%%%%%%%
%%%%%%%%

\subsection{FRR violation spectrum for slow observables}

Consider a special conjugate variable $\mathcal{Q}_{p_k}^s=\mathcal{Q}_p^s$ that is uniform within the same coarse-grained state.  It defines a velocity observable $\dot{Q}^s(t)=\frac{d}{dt}\mathcal{Q}_{p_t}$ which is non-zero only when a slow transition to a neighboring coarse-grained state takes place,  thus classified as a slow observable and emphasized by the superscript $s$.  Below,  we study the asymptotic behavior of its FRR violation spectrum for $\epsilon\to 0$.  The main result has   already been announced  in our previous paper~\cite{Wang2016entropy}, especially in its supplemental material.  Here, we provide more details.

First,  we also estimate the magnitude of the corresponding projection coefficients $\alpha_j^s$, $\beta_j^s$, and $\phi_j^s$ in the limit $\epsilon\to 0$.   The projection on slow modes gives a constant  value, i.e.,  ($j\le K$)
\begin{equation}
\alpha_j^s\sim 1,\quad \beta_j^s\sim 1,
\end{equation}
which is not surprising  according to their definition.  However, the projection on fast modes  is vanishingly small, i.e.,  ($j>K$)
\begin{equation}
\alpha_j^s\sim \epsilon,\quad \beta_j^s\sim \epsilon.
\label{eq:fast-mode-slow-ob}
\end{equation}
This is due to localization of the fast modes within certain coarse-grained states,  and can be verified by using Eq.~(\ref{eq:orthogo}) and Eq.~(\ref{eq:fast-mode-corresp}).   The slow observable $\dot{Q}^s$ is insensitive to flux fluctuation within coarse-grained states,  which gives $B_k^p=0$ in Eq.~(\ref{eq:B-split}).  Therefore,  we always have 
\begin{equation}
\quad \phi_j^s\sim 1,
\end{equation}
whether it is projected on the slow or fast modes.  The magnitude of these the projection coefficients are listed in FIG.~\ref{fig:asymp}.

Intuitively,  the correlation and response of such a slow observable is pretty well described also  on the coarse-grained level in terms of the effective Markov process $\wh{M}$, which involves another set of projection coefficients   $\wh{\alpha}_j^s\equiv\sum_p \wh{x}_j(p)\mathcal{Q}_p^s$,  $\wh{\beta}_j^s\equiv\sum_p \wh{y}_j(p)\mathcal{Q}_p^s\wh{P}_p^{ss}$, and $\wh{\phi}_j^s\equiv \sum_p \wh{y}_j(p)\wh{B}_p$.  Here,  $\wh{B}_p$ is the flux fluctuation defined for $\wh{M}$,  in the same spirit as $B_n$ for $M$ in Eq.~(\ref{eq:Bn-main}),  which satisfies $\wh{B}_p=\sum_k B_{p_k}^{(1)}+O(\epsilon)$ due to the stationary distribution   $P^{ss}_{p_k}=\wh{P}^{ss}_pP(k|p)+O(\epsilon)$.  According to  the  mapping relations for slow modes in Eq.~(\ref{eq:slow-mode-corresp}),    the descriptions at the two levels are related by ($j\le K$)
\begin{subequations}\label{eq:mapping-Qs-slow}
\begin{eqnarray}
 \alpha_j^s& =\wh{\alpha}_j^s+O(\epsilon)\\
  \beta_j^s&= \wh{\beta}_j^s+O(\epsilon)\\
  \phi_j^s&=  \wh{\phi}_j^s+O(\epsilon),
\end{eqnarray}
\end{subequations}
which are exactly the same in terms of  the leading order of the slow modes.  This justifies the validity of coarse-graining if we are interested in this slow observable. 

\begin{figure}
\centering
\includegraphics[width=8cm]{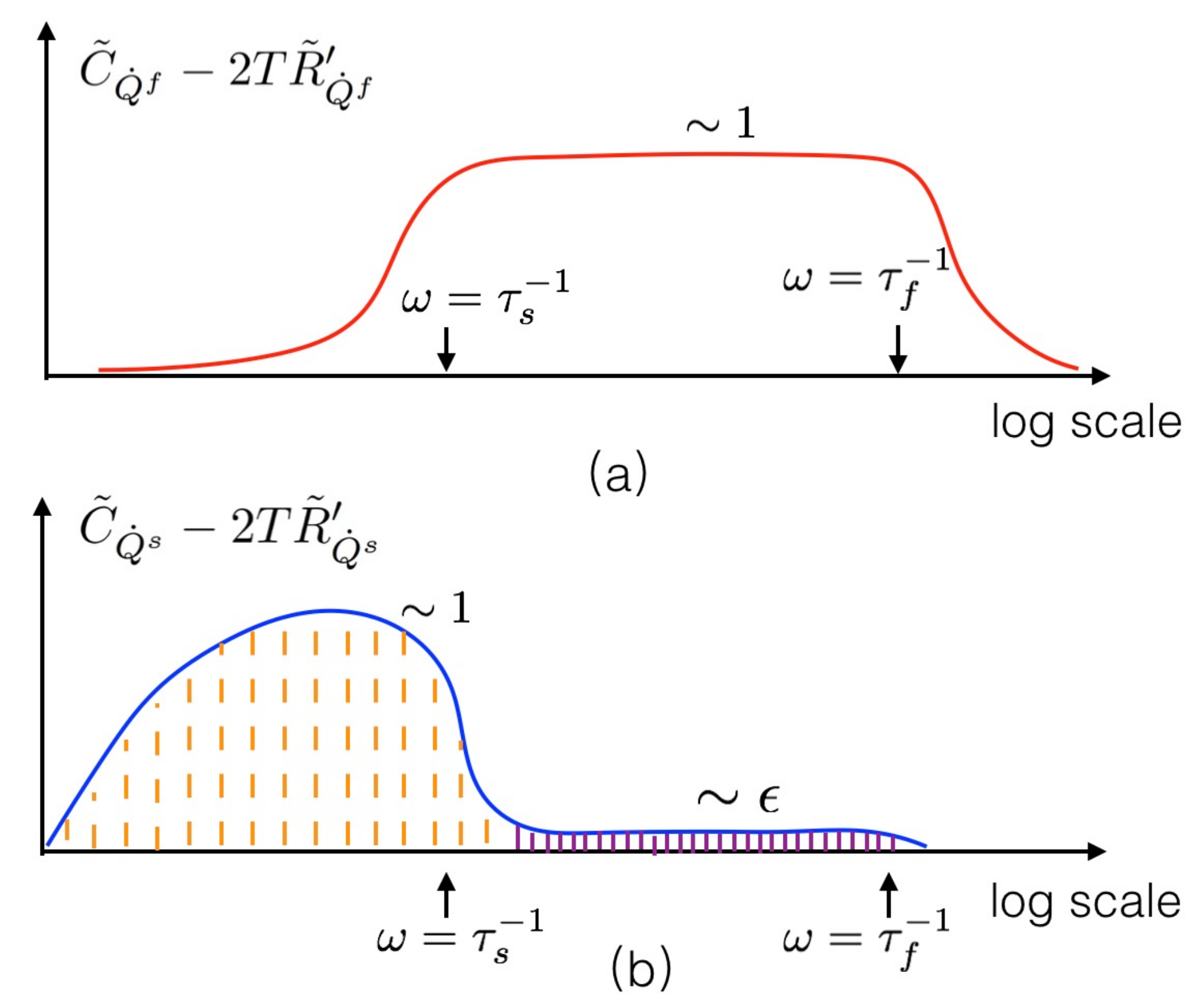}
\caption{(a) Illustration of FRR violation spectrum for a fast observable,  assumping that detailed balance is satisifed within each coarse-grained state. This is equivalent to a illustration of $V_f(\omega)$,  which is related to hidden entropy production.  (b) Illustration of FRR violation spectrum for a slow observable.  The violation in the low frequency region,  shaded by orange dash line,  is due to the broken of detailed balance at the level of effective dynamics,  and the small violation at the intermediate frequency region is related to hidden entropy production.  }
\label{fig:illustration_FRRviolation}
\end{figure}

   The FRR violation spectrum of this slow observable can be split into the contribution from this coarse-grained description  [Eq.~(\ref{eq:velo-FRR-vio})]  and a  correction from the underlying fast processes, i.e., 
 \begin{equation}
 \tilde{C}_{\dot{Q}^s}-2T\tilde{R}'_{\dot{Q}^s}=2\sum_{j=2}^K \widehat{\alpha}_j^s \frac{T\widehat{\phi}_j^s-\widehat{\beta}_j\widehat{\lambda}_j}{1+(\omega/\widehat{\lambda}_j)^2}+\epsilon V_s(\omega).
 \label{eq:velo-FRR-vio-slow}
 \end{equation}
The correction term $ \epsilon V_s(\omega)$ is given by 
\[
\epsilon V_s(\omega)=2\sum_{j\ge 2} \alpha_j^s\frac{T\phi_j^s-\beta_j^s\lambda_j}{1+(\omega/\lambda_j)^2}-2\sum_{j=2}^K \wh{\alpha}_j\frac{T\wh{\phi}_j-\wh{\beta}_j\wh{\lambda}_j}{1+(\omega/\wh{\lambda}_j)^2}.
\]
With the mapping relations [Eq.~(\ref{eq:mapping-Qs-slow})] for the slow modes and the magnitude estimation for the fast modes [FIG.~\ref{fig:asymp}],  we find that $\epsilon V_s(\omega)$ is of order $\epsilon$.  Similar to $V_f(\omega)$, $V_s(\omega)$ also vanishes for both $\omega\gg\lambda_N$ and $\omega\ll \lambda_2$.  In the intermediate frequency region $\tau_s^{-1}\ll\omega\ll \tau_f^{-1}$, only the fast modes contributes, i.e.,  
\begin{equation}
V_s\simeq \frac{2}{\epsilon} \sum_{j>K}\alpha_j^s (T\phi_j^s-\lambda_j\beta_j^s)\xrightarrow{\epsilon\to 0} \text{const},
\label{eq:slow-quasi-sumrule}
\end{equation}
where the limit is obtained by using the magnitude estimation for the fast modes [FIG.~\ref{fig:asymp}].  Alternatively, we may express this plateau in terms of the slow modes by using the sum rule [Eq.~(\ref{eq:sumrule-general})], i.e.,
\begin{equation}
V_s\simeq \frac{2}{\epsilon} \sum_{j=2}^K\alpha_j^s (\lambda_j\beta_j^s-T\phi_j^s)\xrightarrow{\epsilon\to 0} \text{const}.
\label{eq:slow-quasi-sumrule}
\end{equation}
The mapping relations [Eq.~(\ref{eq:mapping-Qs-slow})] and the sum rule for the coarse-grained system, i.e.,  $\sum_{j=2}^K \wh{\alpha}_j(\wh{\lambda}_j\wh{\beta}_j-T\wh{\phi}_j)=0$,  guarantees this finite limit.   See FIG.~\ref{fig:illustration_FRRviolation}(b) for an illustration of $ \tilde{C}_{\dot{Q}^s}-2T\tilde{R}'_{\dot{Q}^s}$.

The FRR violation integral splits into two terms 
\begin{equation}
\Delta_{Q^s}=\wh{\Delta}_{Q^s}+\Delta_{Q^s}^{(1)},
\end{equation}
where the leading term $\wh{\Delta}_{Q^s}$ comes from the effective system $\wh{M}$ and $\Delta_{Q^s}^{(1)}$ is the correction term from the integral of $\epsilon V_s(\omega)$.  Although $\epsilon V_s(\omega)$ is small,  it extends to the high frequency cutoff $1/\tau_f\sim \epsilon^{-1}$,  which makes the integral $\Delta_{Q^s}^{(1)}\sim 1$,  comparable to the leading term $\wh{\Delta}_{Q^s}$.

\section{Discussion}
\label{sect:discussion}

\subsection{Connection between the FRR violation plateau and hidden entropy production}

It is shown by Esposito that steady state entropy production 
\begin{equation}
\sigma\equiv  \sum_{p_k, q_l} P_{p_k}^{ss} w_{p_k}^{q_l} \ln \frac{P_{p_k}^{ss} w_{p_k}^{q_l}}{P_{q_l}^{ss} w_{q_l}^{p_k}}
\end{equation}
 for a Markov system with time scale separation ($\epsilon\to 0$) can be splitted into three parts~\cite{esposito2012stochastic}:  contribution from the coarse-grained dynamics
\begin{equation}
\sigma_1=\sum_{p,q} \wh{w}_p^q \wh{P}^{ss}_p \ln\frac{\wh{w}_p^q \wh{P}^{ss}_p}{\wh{w}_q^p \wh{P}^{ss}_q} ;
\end{equation}
that from the microscopic transition within the same coarse-grained state
\begin{equation}
\sigma_2=\frac{1}{\epsilon} \sum_{p}\wh{P}^{ss}_p\sum_{k,l} P(k|p) w_k^l(p)\ln \frac{P(k|p)w_k^l(p)}{P(l|p)w_l^k(p)};
\end{equation}
and that from the coupling between fast and slow transitions 
\begin{eqnarray}
\sigma_3&\equiv & \sigma-\sigma_1-\sigma_2\nonumber\\
&=&\sum_{p,q}  \wh{w}_p^q \wh{P}^{ss}_p \sum_{k,l} f_{p_k}^{q_l}P(k|p)\ln \frac{f_{p_k}^{q_l}P(k|p)}{f_{q_l}^{p_k} P(l|q)},
\end{eqnarray}
where $f_{p_k}^{q_l}\equiv w_{p_k}^{q_l}/\wh{w}_p^q$ is the conditional transition rate between microscopic state $p_k$ and $q_l$ given that transition between coarse-grained state $p$ and $q$ is already observed.  All these three contributions are non-negative.  Furthermore,  $\sigma_2$,  if exists,  would diverge in the limit of timescale separation $\epsilon\to 0$. 

In the steady state,  the total entropy production rate of the system is equivalent to the total heat dissipation rate
\begin{equation}
J_{tot}\equiv  T\sum_{p_k,q_l} P_{p_k}^{ss} w_{p_k}^{q_l} \ln \frac{ w_{p_k}^{q_l}}{ w_{q_l}^{p_k}}.
\end{equation} 
The reason is that $\sigma-J_{tot}$ gives the change rate of the system entropy, i.e.,
\begin{eqnarray*}
\sigma-\frac{1}{T}J_{tot}&=&\sum_{p_k,q_l} P_{p_k}^{ss} w_{p_k}^{q_l} \ln \frac{ P_{p_k}^{ss}}{ P_{q_l}^{ss}}\\
&=&  \sum_{p_k,q_l} [P_{p_k}^{ss} w_{p_k}^{q_l} -P_{q_l}^{ss} w_{q_l}^{p_k} ]\ln P_{p_k}^{ss}
\end{eqnarray*}
which vanishes in the steady state because $\sum_{q_l} [P_{p_k}^{ss} w_{p_k}^{q_l} -P_{q_l}^{ss} w_{q_l}^{p_k}]=0$.  For the coarse-grained system,  its total heat dissipation rate is given by 
\begin{equation}
\wh{J}_{tot}=T\sum_{p,q} \wh{w}_p^q \wh{P}^{ss}_p \ln\frac{\wh{w}_p^q }{\wh{w}_q^p }.
\end{equation}
For a coarse-grained state $p$ that is assumed to be isolated from other coarse-grained states,  its heat dissipation rate is given by 
\begin{equation}
J_{tot}^p=\frac{1}{\epsilon}\sum_{k,l}P(k|p)  w_k^l(p) \ln\frac{w_k^l(p) }{w_l^k(p) }.
\end{equation}
Similar as $\sigma=J_{tot}/T$,  we have 
\begin{equation}
\sigma_1=\frac{1}{T}\wh{J}_{tot},\quad \sigma_2= \frac{1}{T}\sum_{p}\wh{P}^{ss}_p J_{tot}^p. 
\end{equation}
Therefore, 
\begin{equation}
\sigma_3=\frac{1}{T}\left [J_{tot}-\wh{J}_{tot}-\sum_{p}\wh{P}^{ss}_p J_{tot}^p\right]. 
\end{equation}

The above relations provide a chance to evaluate $\sigma_1$, $\sigma_2$ and $\sigma_3$ by quantifying the dissipation rates through FRR violation spectrum.   However,  a proper observable usually only captures part of the dissipation rate,  as shown in Eq.~(\ref{eq:GHS}).  To capture the total dissipation rate requires very careful design of a set of ``orthogonal'' observables,  with each taking care of dissipation rate from a subset of transitions in the network that are both complementary and non-overlapping.  This also requires the application of the generalized Harada-Sasa equality,  which is only approximately true for Markov jumping systems with proper transition rates as discussed before.   Assuming that all these obstacles can be overcome,  we can use the FRR violation integral for the effective dynamics, i.e., $\wh{\Delta}_{Q^s}$,  to evaluate $\wh{J}_{tot}$,  $\Delta_{Q^f}^p$ for the fast dynamics within a coarse-grained state $p$ to estimate $J_{tot}^p$,   and the correction terms from the violation plateau, i.e., $\Delta_{Q^s}^{(1)}$ and $\Delta_{Q^f}^{(1)}$ will contain information about  $\sigma_3$.  

Now,  we assume that detailed balance is satisfied within each coarse-grained state, which implies that $\sigma_2=0$ and $J_{tot}^p=0$.  In this case,  the potential FRR violation for slow and fast observables are illustrated in FIG.~\ref{fig:illustration_FRRviolation}.  In FIG.~\ref{fig:illustration_FRRviolation}(b) for the slow observables,   one can clearly distinguish the orange area that contributes to $\sigma_1$ and  the purple area that contributes to $\sigma_3$.  The plateau in FIG.~\ref{fig:illustration_FRRviolation}(a) only contributes to $\sigma_3$.  Therefore, $\sigma_1$ comes from the FRR violation in the low frequency region $\omega \sim \tau_s^{-1}$, and $\sigma_3$ comes from the FRR violation in the high frequency region $\omega\sim \tau_f^{-1}$ (note that this pletau is plotted in a log scale and its integral is actually dominated by the region $\omega\sim \tau_f^{-1}$). Then,  it is possible to quantify $\sigma_1$ and $\sigma_3$ only from measurable quantities of the original system.  This will be illustrated later through an Markov jumping example.

Below,  under the assumption that $\sigma_2=0$,   we connect $\sigma_1$ and $\sigma_3$ with FRR violation spectrum for  the $N_0$-component Langevin equation~(\ref{eq:general-Langevin}).  For this system,  not only the Harada-Sasa equality is applicable,  the complete set of orthogonal observables are also well-defined,  which is $\dot{x}_j$ for $j=1,2,\cdots, N_0$.   We assume that the variables are indexed in such a way that $\gamma_j\sim 1$ for $j\le K_0$ and $\gamma_j\sim \epsilon $ for $j>K_0$.   This means that  $x_j$ is a slow variable for $j\le K_0$,  with $K_0$ the number of slow variables in this system.  We also assume that there is no net drift, i.e., $\langle x_j\rangle=0$.  This multi-component langevin system can be described by  our general Markov system with time scale separation, where the coarse-grained state would be any configuration specified by the slow variables, i.e.,  $\vec{x}^s\equiv (x_1,x_2,\cdots, x_{K_0})$,  and the microscopic state within $\vec{x}^s$ would be  any configuration specified by all the fast variables.  
 With these assumptions,  we have 
\begin{subequations}
\begin{eqnarray}\label{eq:sigma-delta}
\sigma_1&=&\frac{1}{T}\wh{J}_{tot}=\frac{1}{T}\sum_{j=1}^{K_0} \gamma_j \wh{\Delta}_{x_j},\\
\sigma_3&=&\frac{1}{T}\sum_{j=1}^{K_0} \gamma_j \Delta_{x_j}^{(1)}+\frac{1}{T}\sum_{j=K_0+1}^{N_0} \gamma_j \Delta_{x_j}.
\end{eqnarray}
\end{subequations}
Here,  $ \wh{\Delta}_{x_j}$ and $\Delta_{x_j}^{(1)}$, with $j=1,2,\cdots, K_0$,  can be evaluated by the integral over the low and high frequency region of the FRR violation spectrum for this slow observable $\dot{x}_j$, respectively.

For many physical systems,  time scale separation usually implies that not only $\sigma_2=0$ but also $\sigma_3=0$, i.e., no hidden entropy production at all.     An interesting example is the potential switching model for molecular motors~\cite{kawaguchi2014nonequilibrium},  where chemical transition is fast and displacement is slow.   In this scenario,   the total dissipation rate could be extracted by only studying the FRR violation of the slow observables,   as long as the slow transitions satisfy our assumptions that lead to Eq.~(\ref{eq:GHS}).

\subsection{Effective temperature for a fast observable}

Here,  we define  effective temperature  by naively assuming the FRR for all frequency:
\begin{equation}
T_{eff}(\omega; \dot{Q})\equiv \frac{\tilde{C}_{\dot{Q}}(\omega)}{2\tilde{R}_{\dot{Q}}'(\omega)}.
\end{equation}
 Note that this definition should be modified accordingly to apply for a displacement observable $Q(t)$,  so as to give the same result.       In general,  $T_{eff}$ depends both on frequency and the observable considered.   It converges to the bath temperature in the high frequency region $\omega\gg\lambda_N$.  
 
  This concept becomes popular after   Cugliandolo \emph{et al.} apply it to glassy systems in~\cite{LeticiaPREtemperature},  where they  argue that the inverse effective temperature actually controls the direction of heat flow.   For example, $1/T_{eff}(\omega,\dot{Q})<1/T$ implies that heat flows from this degree of freedom $Q(t)$ to the bath.   Although $T_{eff}$ share this  desirable property with the temperature of an equilibrium system,   its value is in general sensitive to the choice of observable~\cite{fielding2002observable}.   Below,  we discuss the property of this definition for our system.  Its physical meaning  will be treated in elsewhere.

 We assume that  $\sigma_2=0$, which implies that $T\phi_j^q=\beta_j^q\lambda^q$ for transitions within the same coarse-grained state.    Then, the  effective temperature of a general fast observable is given by 
 \begin{equation}
 T_{eff}(\omega,\dot{Q}^f)=T+ \frac{V_f(\omega)}{2\tilde{R}'_{\dot{Q}^f}(\omega)}.
 \end{equation} 
 At the frequency region $\omega\gg \tau_f^{-1/2}$,     the response spectrum $\tilde{R}'_{\dot{Q}^f}(\omega)\sim \epsilon^{-1}$,  dominated by the fast modes,   while at the frequency  $\omega\ll \tau_f^{-1/2}$  we have $\tilde{R}_{\dot{Q}^f}(\omega)\sim 1$,  dominated by the slow modes.  Besides, $V_f(\omega)$ changes from zero to a finite value around the frequency $\tau_s^{-1}$. Combining these,   we find that this effective temperature takes constant value in three frequency regions: 
 \begin{equation}
 T_{eff}(\omega,\dot{Q}^f)=\begin{cases}
 T+T_1^f, &\omega\ll \tau_s^{-1}\\
 T+T_2^f,&   \tau_s^{-1}\ll\omega\ll \tau_f^{-1/2}\\
 T,&\omega\gg \tau_f^{-1/2}.
 \end{cases}
 \end{equation}
Here,  $T_1^f$ and $T_2^f$ are two different numbers that  satisfy,  up to leading order, 
\begin{subequations}
 \begin{equation}
T_1^f= \frac{\sum_{j=2}^K \alpha_j^f (\beta_j^f \lambda_j-T\phi_j^f)/\lambda_j^2}{\sum_{j=2}^K \alpha_j^f\phi_j^f/\lambda_j^2}+O(\epsilon),
\label{eq:Teff_fast_low_1}
\end{equation}
 \begin{equation}
T_2^f= \frac{\sum_{j=2}^K \alpha_j^f (\beta_j^f \lambda_j-T\phi_j^f)}{\sum_{j=2}^K \alpha_j^f\phi_j^f}+O(\epsilon).
\label{eq:Teff_fast_low_2}
\end{equation}
\end{subequations}
Note that $T_1^f$ has a similar structure as $T_2^f$, except for a weighting factor $1/\lambda_j^2$ that is ordered in a descending way, i.e., $1/\lambda_2^2\ge 1/\lambda_3^2\ge\cdots$.  Both $T_1^f$ and $T_2^f$ pick up an eigenmode that contributes dominantly.   However,  $T_1^f$ favors a slower eigenmode due to the weighting factor.   Only a eigenmode that violates detailed balance can contribute, i.e., $\beta_j^f \lambda_j\neq T\phi_j^f$.  

Through quite a few non-trivial examples,  we find that 
\begin{equation}
T_1^f\approx T_2^f.
\label{eq:T1T2}
\end{equation}
This implies that both $T_1^f$ and $T_2^f$ pick up the slowest eigenmode.  In other words,   the slowest eigenmode ($j=2$)  contributes more significantly to the violation of detailed balance than other slow modes, i.e.,  for $2<j\le K$, 
\begin{equation}
|\alpha_2^f (\beta_2^f \lambda_2-T\phi_2^f)|>|\alpha_j^f (\beta_j^f \lambda_j-T\phi_j^f)|.
\end{equation}
  This agrees with our intuition that slower modes takes longer time to relax to equilibrium and thus breaks detailed balance more easily when driven out of equilibrium.   Eq.~(\ref{eq:T1T2})  implies that this two-timescale system has only two distinct temperatures at the large and slow timescale respectively,  which are   independent of the time scale separation index $\epsilon$.

\subsection{Effective temperature for a slow observable}
The effect of hidden fast processes on a slow observable can be captured by a renormalized (effective) rate matrix Eq.(\ref{eq:effective-M}) and a fast colored noise with small correlation time $\sim \tau_f$.   We assume that the effective dynamics $\wh{M}$ in the slow time scale $\tau_s$ satisfies detailed balance,  and seek to capture the noise effect by an effective temperature.

The effective temperature for a slow observable $Q^s$ consists of a constant bath temperature $T$ and a frequency-dependent component from the colored noise: 
\begin{equation}
T_{eff}(\omega,\dot{Q}^s)=T+\epsilon \frac{V_s(\omega)}{2\tilde{R}'_{\dot{Q}^s}(\omega)}.
\end{equation}
In general,  $\tilde{R}'_{\dot{Q}^s}(\omega)\sim 1$ at all frequency,  dominated by the slow modes. Besides, $V_s(\omega)\sim 1$ and has two cutoffs at the low frequency $\tau_s^{-1}$ and high frequency $\tau_f^{-1}$, respectively.  Therefore, we find that the effective temperature becomes constant in the low, intermediate, and high frequency region: 
\begin{equation}
T_{eff}(\omega;\dot{Q}^s)=
\begin{cases}
T+\epsilon T_1^s,&\quad \omega\ll \tau_s^{-1}\\
 T+\epsilon T_2^s,&\quad   \tau_s^{-1}\ll  \omega\ll \tau_f^{-1}\\
 T,&\quad \omega\gg \tau_f^{-1}.
\end{cases}
\end{equation}
Here,  $T_1^s$ ($T_2^s$) has a similar structure as $T_1^f$ ($T_2^f$), i.e., 
\begin{equation}
T_1^s=\frac{1}{\epsilon} \frac{\sum_{j=2}^K \alpha_j^s (\beta_j^s \lambda_j-T\phi_j^s)/\lambda_j^2}{\sum_{j=2}^K \alpha_j^s\phi_j^s/\lambda_j^2}+O(\epsilon^2),
\label{eq:Teff_low}
\end{equation}
\begin{equation}
 T_2^s=\frac{1}{\epsilon}\frac{\sum_{j=2}^K \alpha_j^s (\beta_j^s \lambda_j-T\phi_j^s)}{\sum_{j=2}^K \alpha_j^s\phi_j^s}+O(\epsilon^2),
\label{eq:Teff_medium}
\end{equation}
where $\beta_j^s \lambda_j-T\phi_j^s\sim \epsilon$ due to our assumption that the system reach equilibrium effectively in the large time scale region,  which ensures that $T_1^s$ and $T_2^s$ are of order 1.   The slow dynamics is frozen  at the intermediate frequency region $ \tau_s^{-1}\ll  \omega\ll \tau_f^{-1}$, while it evolves to a stationary distribution at the time scale much larger than $\tau_s$.  Since the strength of this colored noise generally depends on the value of the slow variable,   it is generally renormalized into different  noise strength at the intermediate and slow frequency region, respectively.    $T_1^s$ or $ T_2^s$ measures the strength of this noise, and thus strength of the driving from the hidden fast processes.   In the limit that $\epsilon\to 0$,  the effective temperature converges to the bath temperature.

We also find that 
\begin{equation}
T_1^s\approx T_2^s.
\label{eq:T1T2-s}
\end{equation}
The underlying mechanism is similar to that of Eq.~(\ref{eq:T1T2}).   Eq.~(\ref{eq:T1T2-s})  implies that we can roughly parameterize the active noise effect by an constant amplitude in the whole frequency region $\omega \ll\tau_f^{-1}$.  In other words,  the system has the same temperatue with the bath at the  time scale smaller than $\tau_f$,  and a slightly different temperature $T+\epsilon T_1^s$ at the  time scale larger than $\tau_f$.

  %However,  the sum rule Eq.~(\ref{eq:sumrule-general}) of the whole eigenmodes  implies that the fast eigenmodes also break the detailed balance condition in a comparable extent.  

\section{Example:  sensory adaptation network}
\label{sect:adaptation}

\begin{figure}[!h]
\centering
\includegraphics[width=8.5cm]{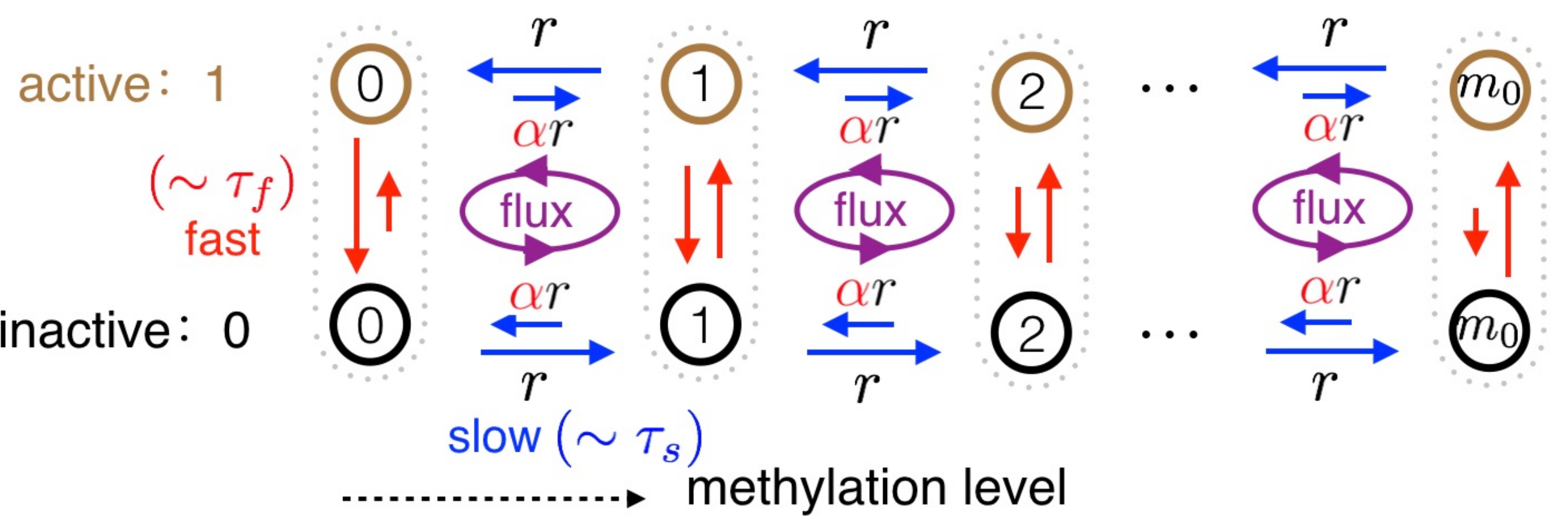}
\caption{Sensory adaptation network for a single membrane receptor in E.coli.  The activity of this receptor $a=1$ in the active state and $0$ when inactive.  Its methylation level $m$ ranges from 0 to $m_0$.  Actually,  $m_0=4$ for this receptor in E.coli.  Here $\alpha\ll 1$ and $\tau_s\gg\tau_f$ to achieve adaptation.  Each methylation level is regarded as a coarse-grained state that contains two microscopic states with different activity. }
\label{fig:adaptation}
\end{figure}

%E.coli has a remarkable membrane receptor  to sense stochastic extracellular ligand binding events,   whose sensory range greatly increased  due to a robust adaptation maintained by the cyclic methylation/demethylation of this receptor on  a slower time scale.  In a Markovian model for this receptor,   its activity and  methylation level together form a state space where stochastic transition takes place between neighboring states.  This system is driven out of equilibrium by cyclic methylation/demethylation events~\cite{lan2012energy, shouwen2015adaptation,Pablo2015Adaptation},  powered by another currency molecule SAM inside E.coli.  This biologically relevant  model is another concrete case of our general setup,  where the methylation level  represents the coarse-grained state within which the activity is the  microscopic state varying on a faster time scale.  

Here,  we study the Markov network illustrated in FIG.~\ref{fig:adaptation},  which describes  sensory adaptation of the membrane receptor in E.coli~\cite{lan2012energy, shouwen2015adaptation,Pablo2015Adaptation}.   This network contains two degrees of freedom:  $a\,(=0,1)$ for the activity of this protein and $m \,(=0,1,\cdots,m_0)$ for its methylation level.  Here,  $a$ changes relatively fast on the time scale $\tau_f$ while  $m$ changes on the slow time scale $\tau_s$.  
 For a fixed $m$,  the activity reaches a local equilibrium distribution $P(1|m)/P(0|m)=\exp(-\Delta S(m))$,  where $\Delta S(m)$ is a linear function given by $\Delta S(m)=e_0 (m_1-m)$.  In general,  the effect of extracellular ligand binding in NESS is captured by a shift of  $m_1$.  We may assume that the activation rate $w_0(m)$ and inactivation rate $w_1(m)$ satisfy 
\begin{subequations}\label{eq:activationRate}
\begin{eqnarray}
w_0(m)&=&\frac{1}{\tau_f}\exp\left(-\frac{\Delta S(m)}{2T}\right),\\
 w_1(m)&=&\frac{1}{\tau_f}\exp\left(\frac{\Delta S(m)}{2T}\right).
\end{eqnarray}
\end{subequations}
  We assume that in the inactive (active) conformation the methylation (demethylation) rate is  $r$,  while  the reverse transition rate is attenuated by a small factor  $\alpha$,   as illustrated  in FIG.~\ref{fig:adaptation}.    The time scale of methylation events is given by $\tau_s=1/r$.   $\alpha\ll r$ is required for high sensory adaptation accuracy in E.coli.    The time scale separation of this system is again captured by $\epsilon\equiv \tau_f/\tau_s$,  which should be small to achieve adaptation.     For more details of this model,  please refer to~\cite{shouwen2015adaptation}.

\begin{figure}[!h]
\centering
\includegraphics[width=8cm]{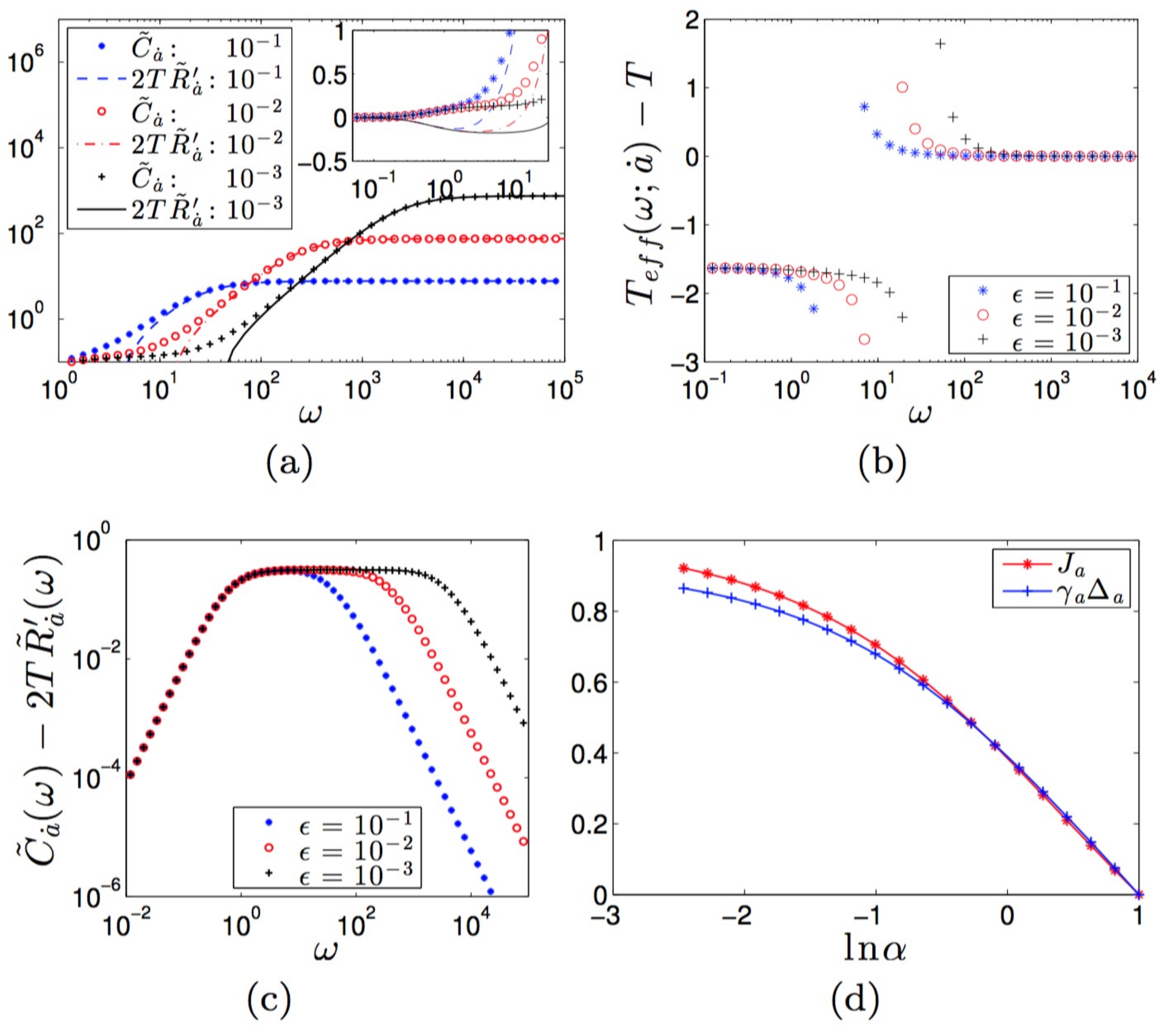}
%\subfigure[]{
%\includegraphics[width=4.3cm]{Adaptation-CorrResp-a}}
%\subfigure[]{
%\includegraphics[width=4cm]{Adaptation_a_Teff}}
%\subfigure[]{
%\includegraphics[width=3.9cm]{AdaptationFDTvio-fre-a}}
%\subfigure[]{
%\includegraphics[width=4.3cm]{a_HS}}
\caption{(a) Correlation and response spectrum for the fast observable $\dot{a}$ at various $\epsilon$.  The inset shows discrepancy of the two spectrum at the low frequency region.  (b) The effective temperature of this observable. (c) The corresponding FRR violation spectrum.   (d) The dissipation rate $J_a$ due to change of only activity and the estimate $\gamma_a\Delta_a$ by assuming Harada-Sasa equality,  with $\gamma_a=1.4\tau_fT$.  Parameter:  $r=1$ (thus, $\epsilon=\tau_f$), $\alpha=0.1$,  $e_0=2$, $T=1$,  $m_0=4$,  and $m_1=m_0/2$. We vary $\tau_f$ to change $\epsilon$.  For (d),  we fix $\tau_f=0.1$ and change $\alpha$ instead.  }
\label{fig:a-adaptation}
\end{figure}

First,  we study  the fast observable $\dot{a}$,  the change rate of the activity.  To obtain its response, we  apply  a perturbation $h$ that increase the activation rate $w_0$ by a factor $\exp(h/2T)$, but decrease the inactivation rate $w_1$ by a factor $\exp(-h/2T)$.    Such a perturbation equivalently modifies  $\Delta S(m)\to \Delta S(m)-h$,  which can be realized by changing the extracellular ligand concentration slightly.  According to the proposed perturbation form Eq.~(\ref{eq:perturbation}),   the activity $a$  is the corresponding conjugate field.  

 FIG.~\ref{fig:a-adaptation}(a) shows the numerically exact   velocity correlation spectrum $\tilde{C}_{\dot{a}}$ and the real part of the response spectrum$\tilde{R}_{\dot{a}}'$ at various $\epsilon$.  We can see that the FRR is approximately satisfied in the high frequency region $\omega\sim \tau_f^{-1}$, but violated  in the low frequency region $\omega\sim \tau_s^{-1}$.  In particular,  the response spectrum $\tilde{R}_{\dot{a}}'$ becomes negative in this low frequency region, while the correlation spectrum $\tilde{C}_{\dot{a}}$ remains positive,  resulting in a negative effective temperature in this low frequency region.  Still,  the inverse effective temperature $1/T_{eff}(\omega,\dot{a})\le 1/T$ at all the frequency, and therefore  the heat also flows  from this degree of freedom $a$ to the bath even at this low frequency region.   This is illustrated in FIG.~\ref{fig:a-adaptation}(b),  where the effective temperature becomes a negative constant for low frequency region $\omega\ll \tau_f^{-1/2}$ and reaches bath temperature for high frequency region $\omega\gg \tau_f^{-1/2}$.  FIG.~\ref{fig:a-adaptation}(c) shows that the corresponding FRR violation spectrum has a plateau in the broad intermediate frequency region $\tau_s^{-1}\ll \omega\ll \tau_f^{-1}$ with an $\epsilon$-independent amplitude,  which agrees with our general analysis.   Although this plateau is of order $1$,  it is much smaller compared with the correlation or response spectrum in the high frequency region,  which is of order $\epsilon^{-1}$.  Indeed,  the frequency-dependence of the effective temperature suggests that the FRR violation is much easier to be detected in the low frequency region.

Next,  we consider the slow observable $\dot{m}$,  the change rate of the methylation level.  To obtain its response,  we use a perturbation $h$ that increases  all the methylation rate (i.e., $r$ for inactive state and $\alpha r$ for active state) by a factor $\exp(h/2T)$,  but decreases all the demethylation rate (i.e., $\alpha r$ for inactive state and $r$ for active state) by a factor $\exp(-h/2T)$.   The conjugate field here is the methylation level $m$.  

\begin{figure}
\centering
\includegraphics[width=8cm]{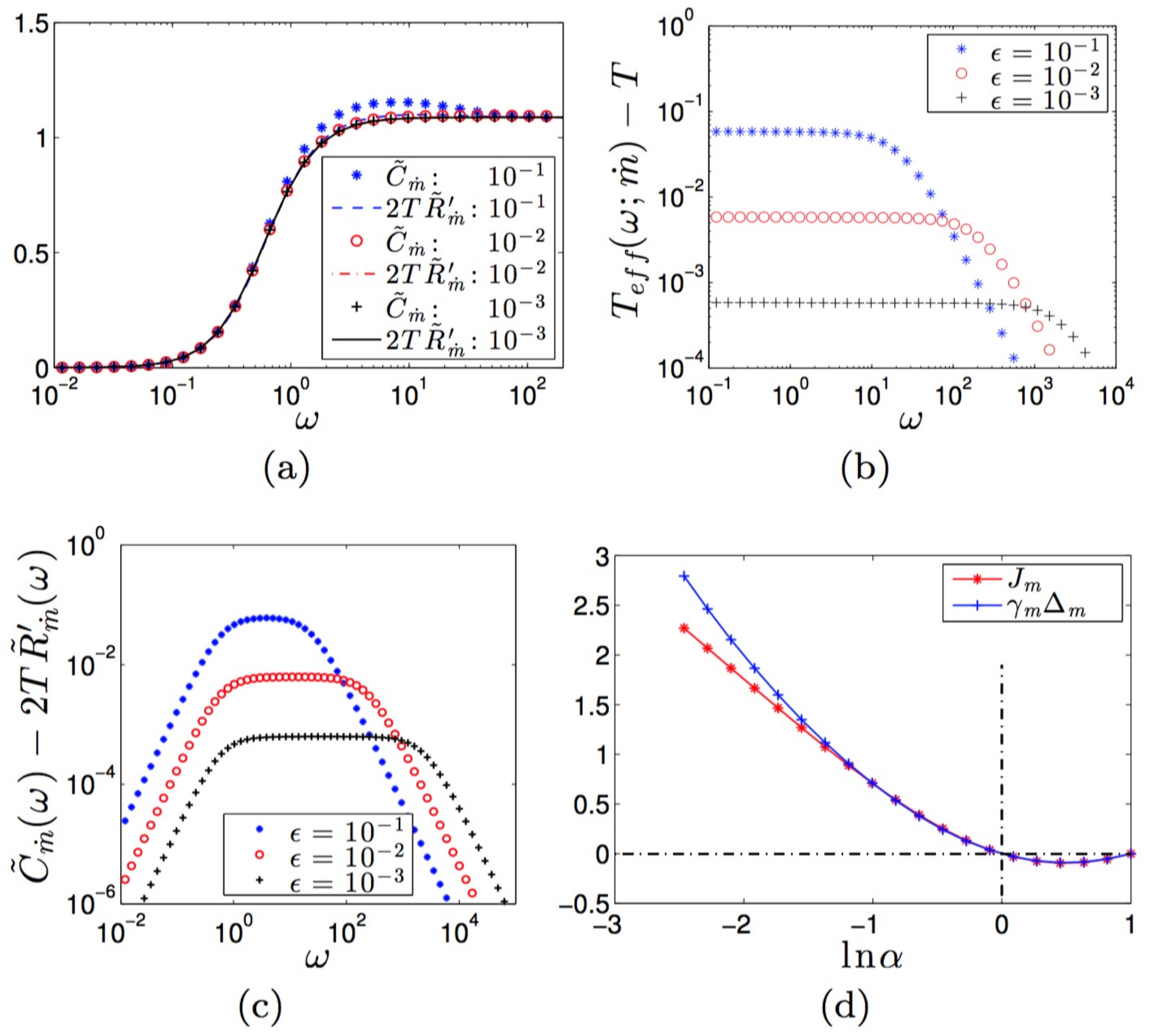}
%\subfigure[]{
%\includegraphics[width=4.2cm]{Adaptation-CorrResp-m}}
%\subfigure[]{
%\includegraphics[width=4cm]{Adaptation_m_Teff}}
%\subfigure[]{
%\includegraphics[width=3.9cm]{AdaptationFDTvio-fre-m}}
%\subfigure[]{
%\includegraphics[width=4.4cm]{m_HS}}
%\includegraphics[width=8.5cm]{adaptation_spectrum}
\caption{  The analysis for the observable $\dot{m}$,  similar to FIG.~\ref{fig:a-adaptation}. Here,     $\gamma_m=T(r\sqrt{\alpha})^{-1}$.  The parameters are the same as those in FIG.~\ref{fig:a-adaptation}.  }
\label{fig:m-adaptation}
\end{figure}

The numerically exact correlation and response spectrum are shown in  FIG.~\ref{fig:m-adaptation}(a).   The violation of FRR appears mainly in the intermediate frequency range and this violation tends to vanish in the limit $\epsilon\to 0$, implying an equilibrium-like dynamics in the time scale separation limit.   The non-equilibrium effect of the hidden fast variable can be captured by the extra effective temperature $T_{eff}-T$,  as shown in  FIG.~\ref{fig:m-adaptation}(b),  which is almost constant in the region $\omega\ll \tau_f^{-1}$ and has a vanishingly small amplitude that scales linearly with $\epsilon$.   The positivity of this extra effective temperature gives rise to a small heat flow from the degree of freedom $m$ to the bath at all frequency.  The FRR violation spectrum serves as another measure of the non-equilibrium effect of the hidden fast processes,  as is shown in FIG.~\ref{fig:m-adaptation}(c).   It has a plateau in the frequency region $\tau_s^{-1}\ll\omega\ll \tau_f^{-1}$ with also a small amplitude  of order $ \epsilon$. This behavior confirms our general analysis.  This FRR violation can be quite difficult to detected experimentally.

 The total dissipation rate of a Markov system is given by 
\begin{equation}
J_{tot}=\frac{1}{2}\sum_{n,n'} (P^{ss}_nw_n^{n'}-P^{ss}_{n'} w_{n'}^n) \ln \frac{w_n^{n'}}{w_{n'}^n}, 
\end{equation}
which is always non-negative in the stationary state.  In our bipartite network,   the total dissipation  can be decomposed into dissipation for change of activity,  denoted as $J_a$,  and for change of methylation level,  denoted as $J_m$.  For example,  
\begin{equation}
J_a=\sum_m \left[ P^{ss}_{1,m}w_1(m)-P^{ss}_{0,m}w_0(m)\right] \ln\frac{w_1(m)}{w_0(m)}.
\label{eq:Ja}
\end{equation}
Then, the total entropy production $\sigma$ in our system satisfies 
\begin{equation}
\sigma=\sigma_3=\frac{1}{T}(J_a+J_m),
\end{equation}
where the first equality holds because our system has an equilibrium effective dynamics for the slow variable $m$, i.e., $\sigma_1=0$, and the fast dynamics for each given methylation level also satisfies detailed balance, i.e., $\sigma_2=0$.  
The result of $J_a$ and $J_m$ is shown in FIG.~\ref{fig:a-adaptation}(d) and FIG.~\ref{fig:m-adaptation}(d),  respectively.   Here,  we fix $\tau_f=0.01$ and change $\alpha$ to tune the dissipation rate of the system.  Note that $J_m>0$ for $\alpha<1$ and $J_m<0$ for $1<\alpha<\exp(1)$,  which implies that $\alpha=1$ is a critical point of the system~\cite{shouwen2015adaptation}.

The generality of Harada-Sasa equality in Langevin systems motivates us to check the following relation between FRR violation and dissipation in this bipartite network: 
\begin{equation}
J_a\approx \gamma_a\Delta_a ,\qquad J_m\approx \gamma_m\Delta_m,
\label{eq:HS-adaptation}
\end{equation}
where  $\Delta_a$ and $\Delta_m$ are the FRR violation integral for the observable $\dot{a}$ and $\dot{m}$, respectively.  And,  $\gamma_a$ and $\gamma_m$ are the corresponding effective friction coefficients,  which is in general not well-defined in Markov systems.   We may still derive $\gamma_m=T(r\sqrt{\alpha})^{-1}$  by focusing on the region that $\alpha$ is close to 1,  where the methylation dynamics is almost diffusive and can be well-approximated by a Langevin equation.   See Supplemental Material in~\cite{Wang2016entropy} for details.   However,  $\gamma_a$ cannot be derived by a similar method because we do not have a Langevin-analogy for this two-state network (with given $m$).  
A naive way to overcome this difficulty is to define $\gamma_a=J_a/\Delta_a$  at a particular set of parameters,  and then use this $\gamma_a$ to check whether $J_a\approx \gamma_a\Delta_a$ holds in a more broad parameter region. In this way,  we take $\gamma_a=1.4\tau_fT$,  determined from $\tau_f=0.1$ and $\alpha=1$. 

\begin{figure}
\centering
\includegraphics[width=8.5cm]{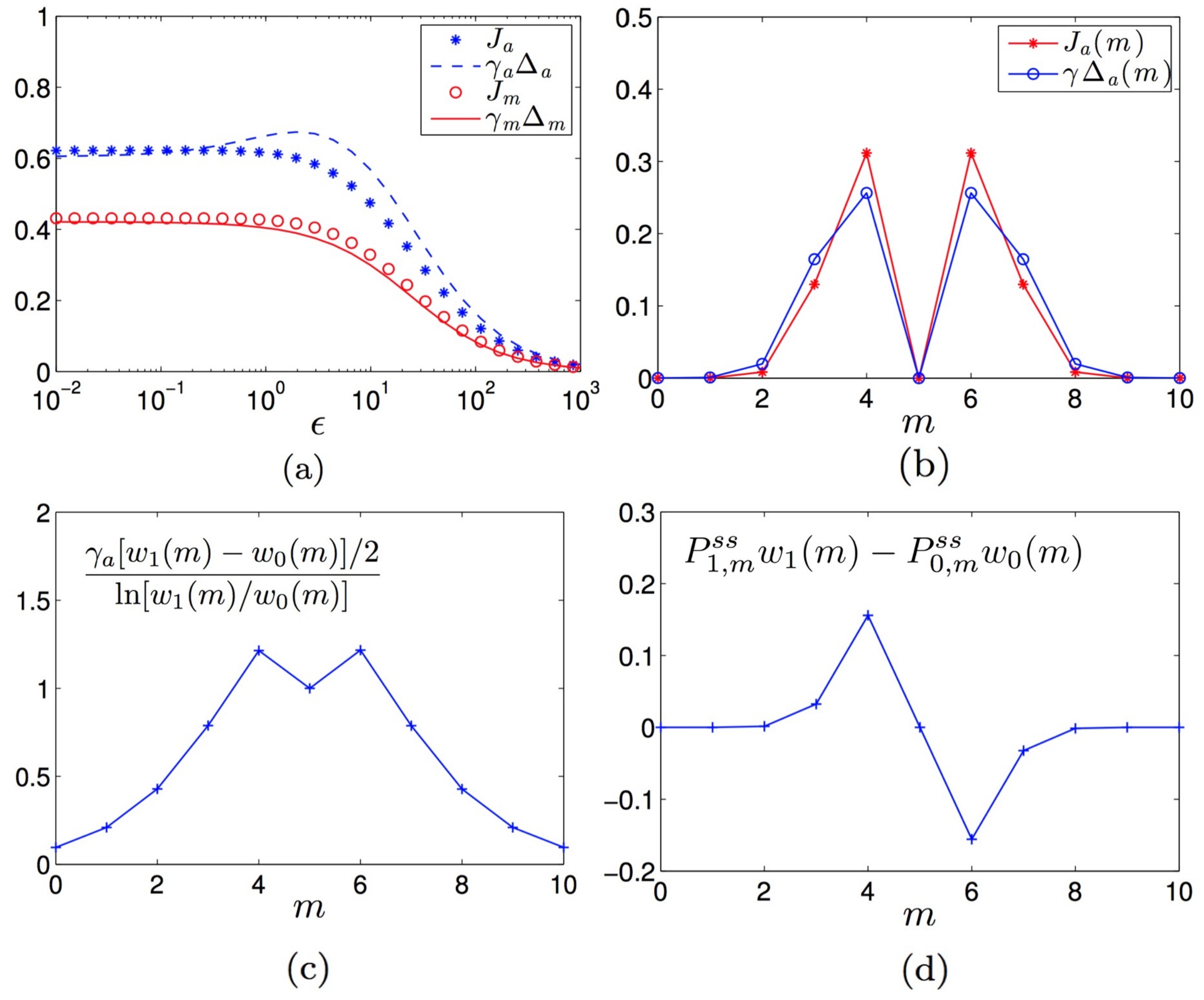}
%\subfigure[]{
%\includegraphics[width=5cm]{Adaptation-Dissipation}}
%\subfigure[]{
%\includegraphics[width=5cm]{flux}}
%\subfigure[]{
%\includegraphics[width=5cm]{HS_comp_1}}
%\subfigure[]{
%\includegraphics[width=5cm]{HS_comp_2}}
%%\subfigure[]{
%%\includegraphics[width=4cm]{Adaptation-Dissipation_log}}
\caption{ (a) Test of generalized Harada-Sasa equality Eq.~(\ref{eq:HS-adaptation}) at different $\epsilon$.  Major violation happens only at $\epsilon \approx 1$.  Here,   $\alpha=0.5$ and $\tau_f$ is varied to change $\epsilon$.  Other parameters are the same as FIG.~\ref{fig:a-adaptation}.  (b)  Comparison  between $J_a(m)$ and $\gamma_a\Delta_a(m)$.  (c) Distribution of the ratio between $\gamma_a(w_1(m)-w_0(m))/2$,  an element from $\gamma_a \Delta(m)$,  and    $\ln [w_1(m)/w_0(m)]$,  an element in $J_a(m)$.   (d) Activity flux distribution.   In (b)(c)(d),  the parameters  are $\alpha=0.1$, $m_0=10$ and  $\tau_f=0.1$.  Other parameters are the same as  FIG.~\ref{fig:a-adaptation}. }
\label{fig:HS-adaptation-2}
\end{figure}

 With these two effective friction coefficients,  we numerically compare $J_x$ and $\gamma_x\Delta_x$ ($x=a,m)$ in FIG.~\ref{fig:a-adaptation}(d) and FIG.~\ref{fig:m-adaptation}(d),  by fixing $\tau_f$ and varying $\alpha$.  Both agree very well in the region $\exp(-1)\le \alpha\le \exp(1)$,  which is very non-trivial because the system displays qualitatively very different behavior for $\alpha>1$ (no adapation with strong boundary effect from $m=0,m_0$) and for $\alpha<1$ (adaptation with negligible boundary effect from $m=0,m_0$)~\cite{shouwen2015adaptation}.  These relations also hold at various $\epsilon$,  as shown  In FIG.~\ref{fig:HS-adaptation-2}.   The Harada-Sasa equality holds approximately for the observable $\dot{m}$ because the methylation dynamics satisfies the assumptions underlying our generalized Harada-Sasa equality~(\ref{eq:GHS}), in particular, a relatively diffusive transition rate.  However,  the validity of $J_a\approx \gamma_a \Delta_a$  would be more difficult to understand,  because this observable has only two states (i.e., $a=0,1$) and the two state dynamics is definitely not quite diffusive since $\Delta S(m)$ varies broadly.  
 
 Below,  we investigate in detail how the Harada-Sasa equality works for the observable $\dot{a}$.   According to Eq.~(\ref {eq:GHSE-2}) and Eq.~(\ref{eq:evaluation}),   we have 
 \begin{equation}
 \Delta_a=\sum_m\left[ P^{ss}_{1,m}w_1(m)-P^{ss}_{0,m}w_0(m)\right] \frac{w_1(m)-w_0(m)}{2}.
 \label{eq:gamma_aDetla_a}
 \end{equation}
 We define $\Delta_a(m)$ to be the contribution of $m$-th methylation level to $\Delta_a$,  and therefore $\Delta_a=\sum_m \Delta_a(m)$.  Similarly,  we define $J_a(m)$ to be the contribution of $m$-th methylation level to $J_a$ in Eq.~(\ref{eq:Ja}),  which satisfies $J_a=\sum_m J_a(m)$.  FIG.~\ref{fig:HS-adaptation-2}(b) shows that $J_a(m)\approx \gamma_a\Delta_a(m)$,  which is necessary to explain why $J_a\approx \gamma_a\Delta_a$  works over such a broad parameter region.   Further analysis reveals that   $\gamma_a(w_1(m)-w_0(m))/2\approx \ln [w_1(m)/w_0(m)]$ holds only for a narrow methylation region where the dissipation $|\Delta S(m)|$ is relatively small,  as shown in FIG.~\ref{fig:HS-adaptation-2}(c).    Fortunately,  the activity flux $[P^{ss}_{1,m}w_1(m)-P^{ss}_{0,m}w_0(m)]$  is significant only around the region where $|\Delta S(m)|$ is also relatively small,  as shown in FIG.~\ref{fig:HS-adaptation-2}(d).    This explains why $J_a(m)\approx \gamma_a\Delta_a(m)$ holds.  
 
 % For this adaptation model,  we have checked that,  under various conditions,  the activity flux is always dominant around the region where $|\Delta S(m)|$ is relatively small.

%  and therefore $J_a(m)\approx \gamma_a\Delta_a(m)$ is quite non-trivial. 
%
%  Note that   the Harada-Sasa equality has been derived only for \emph{diffusive}  Langevin dynamics (i.e., the noise plays dominant role at the short time scale) with an \emph{infinite} state space.   Therefore,  the result that $J_a\approx \gamma_a \Delta_a$ is particularly surprising since this direction 
% 
% Our results suggest that the Harada-Sasa equality might have a wider applicability than Langevin dynamics. 

\section{Conclusion}
\label{sect:conclusion}
Here,  we have made a systematic analysis for a general Markov system with two time scales, focusing on the FRR violation spectrum.  The two characteristic time scales divide the frequency region into three domains:  the low, intermediate, and high frequency region.  Even assuming that the fast processes satisfy detailed balance,   the FRR violation for either a slow or a fast observable is characterized by  a plateau in the intermediate frequency region.  Generically,  this plateau implies a finite hidden entropy production rate that results from coupling between slow and fast processes.  This connection is formulated   precisely for general Langevin systems of two time scales.  A very interesting Markov jumping system motivated from sensory adaptation in E.coli also supports this connection.   To quantify hidden entropy production from FRR violation spectrum,  we need to properly choose a complete set of  orthogonal observables that capture all independent channels of dissiaption,  and measure the FRR violation plateau for each of them.

%An exact connection can be made for 

%With the help of the Harada-Sasa equality,  we find  for general Langevin systems of two time scales that the low frequency FRR violation gives the entropy production of the coarse-grained system,  while the plateau gives the hidden entropy production, a partition proposed previously by Esposito~\cite{esposito2012stochastic}.   This connection might be  OK for discrete Markov systems. 

%In our recent paper,  we have proposed that we could probe hidden dissipation by just looking a slow observable with sufficient temporal resolution~\cite{Wang2016entropy}.   In general, the trajectory a specific observable gives FRR violation,  thus dissipation,  for only this channel. Therefore,   to capture the total dissipation (or entropy production) rate of the system,  we still need to monitor all the independent degrees of freedom,  although certain channels may contribute dominantly.  

% By making use of the Harada-Sasa equality,  we show for Langevin dynamics that the low frequency violation gives the entropy production of the coarse-grained system,  and the high frequency violation for the entropy production within the same coarse-grained state.  In particular,  the unexpected FRR violation in the intermediate frequency region, i.e., the plateau,  reflects the entropy production due to the coupling between slow and fast processes.  Such a correspondence seems to be robust,  and valid even for discrete Markov systems.   Therefore,  we have connected our work with previous studies on hidden entropy production,  by providing a new perspective.  

We have also studied a different measure of non-equilibrium dynamics:  effective temperature.  For a NESS only driven by the coupled motion between fast and slow processes,   we find that the effective temperature for a fast observable approaches the room temperature in the high frequency region, while it significantly deviates from the room temperature  in the low frequency region.  This two-temperature two-time scale scenario is similar with glassy systems~\cite{LeticiaPREtemperature, Kurchan2005nonEQ}.   However,  the effective temperature for a slow observable approaches the bath temperature throughout all frequency region in the timescale separation limit, which is consistent with the emergence of an effective potential landscape.   The tiny deviation of order $\epsilon $ from the bath temperature appears at the low and intermediate frequency region,  which is crucial to explain the finite dissipation rate of the slow variable.   This extra deviation could be modeled by an extra fast noise,  which would then  capture the feature of a finite dissipation rate.   The above results suggest that it is much easier to probe hidden entropy production by measuring the low frequency violation of the FRR for a fast observable.

% $T+\epsilon T_1^s$ in both the intermediate and low frequency region,  which approaches the bath temperature in the time scale separation limit $\epsilon\to 0$.   Therefore,  a tiny deviation from the bath temperature implies a finite dissipation rate underlying the apparant equilibrium slow dynamics.  This implies that we may model  a slow variable by an effective dynamics from coarse-graining plus a  noise that has an amplitude $T_1^s$ and a corresponding fast time scale.   

 The Harada-Sasa equality was originally derived only for  Langevin dynamics  with an infinite state space.  Here,  we also present a systematic discussion on the applicability of this equality to general Markov jumping systems.  We find that  the generalized Harada-Sasa equality not only requires a relatively \emph{diffusive}  transition along the direction of the observable,  the prefactor of the transiton rate,  which quantifies its time scale,  should also be homogeneous along this direction.    Here,  a transition is called to be diffusive when it produces only a small amount of entropy in the medium.   These requirements allow to lump all the transitions in this direction together,  and therefore to access their dissipation rate by only monitoring the stochastic evolution projected along this direction.   Other details such as the system size are not relevant.   In some cases,  although the relevant transitions are not always diffusive,  the transitions that are more diffusive may take place much more frequently,  which may again restore this equality,  as supported by our sensory adaptation model.    This generalized Harada-Sasa equality can be very useful to measure the total entropy production rate for a system with time scale separation,  because the fast processes usually reach equilibrium and that the phenomenon of hidden entropy production is not so common in physical systems.   
 
 %Therefore,  the result that $J_a\approx \gamma_a \Delta_a$ is particularly surprising since this direction   Although the Harada-Sasa equality is not applicable to general discrete Markov systems,  surprisingly,  it remains valid approximately for our bipartite network model  on sensory adaptation.    This suggests that the Harada-Sasa equality may find more application to systems with non-diffusive dynamics (i.e.,  each transition is strongly biased due to large dissipation) and with finite state space.  
 
 Our Markov system assumes a finite state space,  which forbids a net drift of any observables. However,  many interesting periodic systems are able to reach NESS and at the same time have a drifting motion,  say molecular motors.  For such systems,  another key difference is that the correlation and response spectrum will not vanish at the low frequency limit,  and consequently FRR violation at the low frequency limit is also possible.  However,  the other features identified in our current paper will remain unchanged.   This discussion will be presented elsewhere. 
 
   In the near future,  it would be more interesting to apply our general framework to interesting systems with hidden entropy production,  say  (possibly) inefficiently molecular motors~\cite{toyabe2015single}, active cytoskeletal networks~\cite{Mizuno2007cytoskeletonNetwork},  and  repulsive self-propelled particles~\cite{Tailleur2008RunTumble},  where gas and liquid phase coexist. 

\section*{Acknowledgements}

 The authors thank  M. Esposito and Y.  Nakayama  for fruitful discussions and comments.  
The work was supported in part by the NSFC under Grant No. U1430237 and by the Research Grants Council of the Hong Kong Special Administrative Region (HKSAR) under Grant No. 12301514.  It was also supported by KAKENHI (Nos. 25103002 and 26610115), and by the JSPS Core-to-Core program ``Non-equilibrium dynamics of soft-matter and information''.
 
% We know quite a few examples where a non-equilibrium system is mapped to a equilibrium system when the large scale dynamics is interested,  
 
 %To generalize our discussion to such periodic systems with an infinite state space,  we need first to decompose the correlation and response spectrum along the eigenspace of such a system.  This is not a trivial task since the ground state is degenerate for this kind of system. 

%\bibliography{sensoryNetwork2}


%
%\def\theequation{S\arabic{equation}}
%\makeatletter
%\@addtoreset{equation}{section}
%\makeatother
%
%\setcounter{equation}{0}
%\setcounter{section}{0}
%%\preprint
%\newpage
%\section{Supplemental Material}

\onecolumngrid
\appendix

\section{Real part of the response spectrum}
\label{app:realPart}
Now,  we prove that the following summation 
\[\sum_{j=2}^N \alpha_j\phi_j \left[ 1-\frac{1}{1+(\omega/\lambda_j)^2} \right]\]
gives a real function over frequency domain,  although each component can be a complex function.  

 For the first quantity,  inserting the definition of the projection coefficients $\alpha_j$ and $\phi_j$,  we obtain 
  \begin{eqnarray}
\sum_{j=2}^N \alpha_j\phi_j \left[ 1-\frac{1}{1+(\omega/\lambda_j)^2} \right] &=& \sum_{j=1}^N \alpha_j\phi_j \left[ 1-\frac{1}{1+(\omega/\lambda_j)^2} \right]\\
&=& \sum_{n,m}\mathcal{Q}_n\left( \sum_{j=1}^N x_j(n) \Big[1- \frac{\lambda_j^2}{\lambda_j^2+\omega^2}\Big]  y_j(m)\right) B_m .\nonumber\\
   &=& \sum_{n,m}\mathcal{Q}_n\left(1- \frac{ M^2}{M^2+\omega^2} \right)_{nm} B_m ,\nonumber\\
   \end{eqnarray}
  which indeed is a real function ($M$ is a real matrix).  Here, $(\cdot )_{nm}$ takes the entry of the matrix at $n$-th row and $m$-th column.  The above calculation has used a critical  relation 
  \begin{equation}\label{eq:general-sum-j}
   \sum_{j=1}^N x_j(n) f(\lambda_j) y_j(m)=\Big(f(M)\Big)_{nm},
  \end{equation}
  where $f(\cdot)$ is an analytical function.  It can be proved as follows
  \begin{eqnarray*}
  \sum_{j=1}^N x_j(n) f(\lambda_j) y_j(m)&=&\sum_{j=1}^N \sum_{k} \Big(f(M)\Big)_{nk} x_j(k)y_j(m)\\
  &=&  \sum_{k} \Big(f(M)\Big)_{nk} \left[\sum_{j=1}^N x_j(k)y_j(m)\right] \\
    &=&  \sum_{k} \Big(f(M)\Big)_{nk}\delta_{km} 
  \end{eqnarray*}
   Then summation over $k$ gives the desired relation in Eq.~(\ref{eq:general-sum-j}).  
   
Similarly, we can show that 
\[ \sum_{j=2}^N \alpha_j\phi_j \left[ \frac{\omega/\lambda_j}{1+(\omega/\lambda_j)^2 }\right] \]
is a real function,  and thus the real part of the response spectrum is given by 
\begin{equation}
\tilde{R}_{\dot{Q}}'(\omega)=\sum_{j=2}^N \alpha_j\phi_j \left[ 1-\frac{1}{1+(\omega/\lambda_j)^2} \right].
\end{equation}
This justifies the violation spectrum in Eq.~(\ref{eq:velo-FRR-vio}).

\section{The sum rule}
\label{app:sum-rule}  
  
Here,  we prove the sum rule in Eq.~(\ref{eq:sumrule-general}).    Using the definition $\beta_j\equiv \sum_n y_j(n)P^{ss}_n \mathcal{Q}_n$ and the characteristic equation$\lambda_jy_j(m)=- \sum_n  y_j(n)M_{nm}$,  we find that \begin{equation}
\lambda_j \beta_j= \sum_{n,m}  y_j(n)  \Big[w_n^mP_n^{ss}\mathcal{Q}_n-w_m^nP_m^{ss}\mathcal{Q}_m \Big].
\label{eq:lambda-beta}
\end{equation}
Combined with  $\phi_j\equiv \sum_n B_n y_j(n)$ and Eq.~(\ref{eq:Bn-main}), we obtain 
\begin{equation}
\lambda_j\beta_j-T\phi_j =\sum_{n,m}y_j(n)\left( w_n^mP_n^{ss}-w_m^nP_m^{ss} \right)\frac{\mathcal{Q}_n+\mathcal{Q}_m}{2}. \label{eq:lambda-beta-phi}
\end{equation}
Now,  we add another component $\alpha_j=\sum_{n'} x_j(n')\mathcal{Q}_{n'}$,  and check the  summation over all eigenmodes:  $\sum_{j} \alpha_j (\lambda_j \beta_j-T\phi_j)$.   Noting the completeness relations $ \sum_j x_j(n')y_j(n)=\delta_{n',n}$,  we have 
%\begin{eqnarray}
%&&\sum_{j=1}^N \alpha_j (\lambda_j \beta_j-T\phi_j)=\sum_{n,m}\mathcal{Q}_{n} \left( w_n^mP_n^{ss}-w_m^nP_m^{ss} \right)\mathcal{Q}_n/2\nonumber\\
%&&\qquad \qquad  +\frac{1}{2}\sum_{n,m}\mathcal{Q}_{n}  w_n^mP_n^{ss}\mathcal{Q}_m-\frac{1}{2}\sum_{n,m}\mathcal{Q}_nw_m^nP_m^{ss} \mathcal{Q}_m\nonumber.
%\end{eqnarray}
\[\sum_{j=1}^N \alpha_j (\lambda_j \beta_j-T\phi_j)=\frac{1}{2}\sum_{n,m}\mathcal{Q}_{n} \left( w_n^mP_n^{ss}-w_m^nP_m^{ss} \right)\mathcal{Q}_n +\frac{1}{2}\sum_{n,m}\mathcal{Q}_{n} ( w_n^mP_n^{ss}-w_m^n P_m^{ss}) \mathcal{Q}_m.\]
On the right hand side (rhs),  the stationary condition $\sum_m [w_n^mP_n^{ss}-w_m^nP_m^{ss}]=0$ demands that   the first term vanishes,  and   the second term cancels the third term due to their equivalence under exchange of $n$ and $m$.  Then,   we obtain 
\[\sum_{j=1}^N \alpha_j (\lambda_j \beta_j-T\phi_j)=0.\]
   Noting  $\lambda_1=0$  and $\phi_1=0$,  we have derived the sum rule in the main text.

\section{Detailed balance in each eigenmode for equilibrium system}
\label{app:FRReigenmode}
 We have considered a symmetric form of perturbation [Eq.(~\ref{eq:perturbation})] in the main text.  Combining detailed balance condition
 \begin{equation}
 w_n^mP^{eq}_n=w_m^n P^{eq}_m
 \label{eq:detailed-balance-state}
 \end{equation}
 and Eq.~(\ref{eq:lambda-beta-phi}) (note that $ss$ should be replaced by $eq$ since we assume that steady state is an equilibrium state),  we can immediate obtain Eq.~(\ref{eq:detailed-eigenmode}),  the detailed balance in each eigenmode. 
  However, Eq.(~\ref{eq:perturbation})  is only a special case of the general form of perturbation:
\begin{equation}
 \frac{\tilde{w}_m^n}{\tilde{w}_n^m}=\frac{w_m^n}{w_n^m}\exp\left(\frac{1}{T} [\mathcal{Q}_n-\mathcal{Q}_m]h\right).
 \label{eq:perturbation-general}
 \end{equation}
Here, we prove  Eq.~(\ref{eq:detailed-eigenmode}) holds  under this general assumption.  

According to Eq.~(\ref{eq:general-Bn}),  the flux fluctuation vector satisfies  
\begin{eqnarray}
B_n^{eq}&\equiv&\lim_{h\to 0}\partial_h \sum_{m}\tilde{M}_{nm}P^{eq}_m\nonumber\\
&=&\lim_{h\to 0} \frac{\partial }{\partial h}\sum_m \left( \tilde{w}_m^n P_m^{eq}-\tilde{w}_n^m P_n^{eq} \right)\nonumber\\
&=&\lim_{h\to 0}\frac{\partial }{\partial h}\left(\sum_m \left[ \frac{\tilde{w}_m^n P_m^{eq}}{\tilde{w}_n^m P_n^{eq} }-1\right]\tilde{w}_n^m P_n^{eq} \right),\label{eq:Bn-mid-1}
\end{eqnarray}
which holds for any form of perturbation.   Using Eq.~(\ref{eq:perturbation-general}) and $P_n^{ss}=P_n^{eq}+O(h)$,  we have 
\begin{eqnarray*}
\frac{\tilde{w}_m^nP_m^{eq}}{\tilde{w}_n^mP_n^{eq}}&=&\frac{w_m^n P_m^{eq}}{w_n^m P_n^{eq}}\exp\left(\frac{1}{T} [\mathcal{Q}_n-\mathcal{Q}_m]h\right)\\
&=& \frac{w_m^n P_m^{eq}}{w_n^m P_n^{eq}}\left(1+\frac{1}{T} [\mathcal{Q}_n-\mathcal{Q}_m]h\right)+O(h^2)\\
&=& 1+\frac{1}{T} [\mathcal{Q}_n-\mathcal{Q}_m]h+O(h^2).
\end{eqnarray*}
To arrive at the last line,   we have used the detailed balance condition Eq.~(\ref{eq:detailed-balance-state}).   Inserting this result into Eq.~(\ref{eq:Bn-mid-1}) and taking the limit $h\to 0$,  we obtain
\begin{equation}
B_n^{eq}= \frac{1}{T} \sum_m w_m^nP_m^{eq} (\mathcal{Q}_n-\mathcal{Q}_m),
\label{eq:Bn_reduced}
\end{equation}
which is exactly the same as the equilibrium form of  Eq.~(\ref{eq:Bn-main}).   Therefore, $B_n^{eq}$ is independent of the specific form of perturbation and consequently  Eq.~(\ref{eq:detailed-eigenmode})  holds for each eigenmode. 

%Imposing  the detailed balance condition also on $\lambda_j\beta_j$ in  Eq.~(\ref{eq:lambda-beta}),  a straightforward comparison between the result $\lambda_j\beta_j^{eq}$ and $\phi_j^{eq}=\sum_n B_n^{eq}y_j(n)$ gives the desired relation $ \lambda_j\beta_j^{eq}=T\phi_j^{eq}$.  

\section{First order correction}
\label{sect:higher-correction}
Now, we consider the first order correction.  For convenience,  we introduce a compact form $(g,f)$ as inner product between state function $g(n)$ and $f(n)$, i.e., $(g,f)\equiv \sum_n g(n)f(n)$.   It is convenient to decompose $x_j^{(1)}$ and $y_j^{(1)}$ to the  eigenspace of $M^{(1)}$, i.e.,
\begin{subequations}\label{eq:xy-component}
\begin{eqnarray}
x^{(1)}_j&=&\sum_{j'}x_{j'}^{(0)}\left(y_{j'}^{(0)},x^{(1)}_j\right), \\
y^{(1)}_j&=&\sum_{j'}y_{j'}^{(0)}\left(y_{j}^{(1)},x^{(0)}_{j'}\right), \\
\end{eqnarray}
\end{subequations}
and then solve each component accordingly.  By subsituting into the $O(1)$ order equation Eq.~(\ref{eq:perturb-0}),  and then projecting the resulting vector equations onto $j'$-th mode again, we easily obtain the first order correction for fast modes ($j>K$), i.e.,
\begin{subequations}
\begin{eqnarray}
x^{(1)}_j&=&\sum_{j'\neq j} x_{j'}^{(0)} \frac{A_{j',j} }{\lambda_{j'}^{(-1)}-\lambda_j^{(-1)}},\\
y^{(1)}_j&=& \sum_{j'\neq j}   y_{j'}^{(0)}\frac{A_{j,j'}}{\lambda_{j'}^{(-1)}-\lambda_j^{(-1)}},
\end{eqnarray}
\end{subequations}
where $A_{j,j'}\equiv \left(y_{j}^{(0)},M^{(0)}x^{(0)}_{j'}\right)$.   The correction of eigenvalue $\lambda^{(0)}$ for $j$-th fast mode is obtained by projecting Eq.~(\ref{eq:perturb-0}) onto $j$-th fast mode, which gives $(j>K)$
\begin{equation}
\lambda^{(0)}_j=-A_{j,j}.
\end{equation}

The treatment for the slow modes ($j\le K$) requires more care.   Substituting Eq.~(\ref{eq:xy-component}) into Eq.~(\ref{eq:perturb-0}), we find that only the components along the fast modes matters,  and we obtain the projection coefficient along the fast modes $(j\le K,\;j'>K)$:
\[ (y_{j'}^{(0)}, x_j^{(1)})=\frac{A_{j',j}}{\lambda_{j'}^{(-1)}}. \]
\[ (y_{j}^{(1)}, x_{j'}^{(0)})=\frac{A_{j,j'}}{\lambda_{j'}^{(-1)}}. \]
The  components along the slow modes can be obtained by considering equation of order $\epsilon$
\begin{subequations}\label{eq:perturb-1}
\begin{eqnarray}
 -\lambda_j^{(-1)}x^{(2)}(p_k)  -\lambda_j^{(0)}x_j^{(1)}(p_k)-\lambda_j^{(1)}x_j^{(0)}(p_k)&=&\sum_l M^{p}_{kl}x^{(2)}_j(p_l)+\sum_{q_l} M^{(0)}_{p_kq_l}x^{(1)}(q_l)\\
-\lambda_j^{(-1)}y^{(2)}(p_k) -\lambda_j^{(0)}y_j^{(1)}(p_k) -\lambda_j^{(1)}y_j^{(0)}(p_k)&=&\sum_l y^{(2)}_j(p_l) M^{p}_{lk}+\sum_{q_l} y^{(1)}(q_l)M^{(0)}_{q_lp_k}.
\end{eqnarray}
\end{subequations}
  Again,  by substitution and projection,  we obtain  for $j,j'\le K$ 
\[(y_{j'}^{(0)},x_j^{(1)})= \frac{1}{\lambda_{j'}^{(0)}-\lambda_j^{(0)}}\sum_{j''>K}A_{j',j''}\left(y_{j''}^{(0)}, x^{(1)}_j\right),  \]
\[(y_{j}^{(1)},x_{j'}^{(0)})= \frac{1}{\lambda_{j'}^{(0)}-\lambda_j^{(0)}}\sum_{j''>K}\left(y_{j}^{(1)}, x_{j''}^{(0)}\right)A_{j'',j'},  \]
which turns out to be generated by first order correction in the fast modes.   The correction of eigenvalue $\lambda^{(1)}_j$ is given by projecting Eq.~(\ref{eq:perturb-1}) onto $j$-th slow mode, i.e., ($j\le K$)
\begin{equation}
\lambda_j^{(1)}=-(y_j^{(0)}, M^{(0)}x_j^{(1)})=-\sum_{j'>K}\frac{A_{j,j'}A_{j',j}}{\lambda_{j'}^{(-1)}},
\end{equation}
where only the fast modes matter.  Inserting these coefficients back in Eq.~(\ref{eq:xy-component}),  we obtain the first order corretion for the  slow modes.

%\section{Correlation and response for an periodic running system}
%Consider the following equation 
%\[ v(t)-v_s=f(x(t))+\xi(t). \]
%For the velocity correlation 
%\[ \langle [v(t)-v_s][v(0)-v_s]\rangle \]
%we may try to use 
%\[v(t)-v_s=G(t)[v(0)-v_s]\]
%Then,  we have 
%\[ \langle [v(t)-v_s][v(0)-v_s]\rangle=\langle G(t)\rangle \langle [v-v_s]^2\rangle \]
%where we have used Markov property of the dynamics such that $G(t)$ ($t>0$) should be independent of $v(0)$,  and the ensemble average can be split into two parts. 

%\section{Effective temperature for different models}
%\begin{figure}[!h]
%\centering
%\subfigure[]{
%\includegraphics[width=4cm]{laserSwitch_linear_Teff}}
%\subfigure[]{
%\includegraphics[width=4cm]{laserSwitch_nonlinear_Teff}}
%\subfigure[]{
%\includegraphics[width=4cm]{Adaptation_nonhomogeous_m_Teff}}
%\subfigure[]{
%\includegraphics[width=4cm]{BrownianRatchet_Teff}}
%\caption{(a) Linear potential switching model; (b) nonlinear potential  switching model with $U_\sigma(x)=k(x+\sigma L/2)^2/2+5k\exp(-(x+\sigma L/2)^4)$ where $\sigma=\pm 1$; (c) original sensory adaptation network;   (d) sensory network modified by inhomogeneous methylation rate; (e) Brownian ratchet}
%\label{fig:effective-temperature}
%\end{figure}

%\bibliography{/Users/wangshouwen/BaiduYun/MyDocument/sensoryNetwork2}
%\bibliography{sensoryNetwork2}
\end{document}